\documentclass[times,tight]{aastex631}
\usepackage{amsfonts,times}
\usepackage{mathrsfs}
\usepackage{bm}
\usepackage{physics}
\usepackage{gensymb}
\usepackage{lipsum}
\usepackage[OT1]{fontenc} 
\usepackage{xcolor}

\let\tablenum\relax
\usepackage{siunitx}

\received{***}
\revised{***}
\accepted{***}

\shortauthors{Songsheng et al.}

\begin{document}

\title{\large \bf Search for Continuous Gravitational Wave Signals in Pulsar Timing Residuals: A New Scalable Approach with Diffusive Nested Sampling}
\correspondingauthor{Jian-Min Wang}
\email{wangjm@ihep.ac.cn}

\author{Yu-Yang Songsheng}
\affil{Key Laboratory for Particle Astrophysics,
Institute of High Energy Physics,
Chinese Academy of Sciences,
19B Yuquan Road, Beijing 100049, China}
\affil{University of Chinese Academy of Sciences,
19A Yuquan Road, Beijing 100049, China}

\author{Yi-Qian Qian}
\affil{MOE Key Laboratory of Fundamental Physical Quantities Measurements, Hubei Key Laboratory of Gravitation and Quantum Physics, PGMF, Department of Astronomy and School of Physics, Huazhong University of Science and Technology, Wuhan 430074, China}

\author{Yan-Rong Li}
\affil{Key Laboratory for Particle Astrophysics,
Institute of High Energy Physics,
Chinese Academy of Sciences,
19B Yuquan Road, Beijing 100049, China}

\author{Pu Du}
\affil{Key Laboratory for Particle Astrophysics,
Institute of High Energy Physics,
Chinese Academy of Sciences,
19B Yuquan Road, Beijing 100049, China}

\author{Jie-Wen Chen}
\affil{MOE Key Laboratory of Fundamental Physical Quantities Measurements, Hubei Key Laboratory of Gravitation and Quantum Physics, PGMF, Department of Astronomy and School of Physics, Huazhong University of Science and Technology, Wuhan 430074, China}

\author{Yan Wang}
\affil{MOE Key Laboratory of Fundamental Physical Quantities Measurements, Hubei Key Laboratory of Gravitation and Quantum Physics, PGMF, Department of Astronomy and School of Physics, Huazhong University of Science and Technology, Wuhan 430074, China}

\author{Soumya D. Mohanty}
\affil{Department of Physics and Astronomy, The University of Texas Rio Grande Valley, One West University Blvd.,
Brownsville, Texas 78520, USA}

\author{Jian-Min Wang}
\affil{Key Laboratory for Particle Astrophysics,
Institute of High Energy Physics,
Chinese Academy of Sciences,
19B Yuquan Road, Beijing 100049, China}
\affil{University of Chinese Academy of Sciences,
19A Yuquan Road, Beijing 100049, China}
\affil{National Astronomical Observatories of China,
Chinese Academy of Sciences,
20A Datun Road, Beijing 100020, China}

\begin{abstract}
Detecting continuous nanohertz gravitational waves (GWs) generated by individual close binaries of supermassive black holes (CB-SMBHs) is one of the primary objectives of pulsar timing arrays (PTAs).
The detection sensitivity is slated to increase significantly as the number of well-timed millisecond pulsars will increase by more than an order of magnitude with the advent of next-generation radio telescopes.
Currently, the Bayesian analysis pipeline using parallel tempering Markov chain Monte Carlo has been applied in multiple studies for CB-SMBH searches, but it may be challenged by the high dimensionality of the parameter space for future large-scale PTAs. 
One solution is to reduce the dimensionality by maximizing or marginalizing over uninformative parameters semi-analytically, but it is not clear whether this approach can be extended to more complex signal models without making overly simplified assumptions.
Recently, the method of diffusive nested (DNest) sampling shown the capability of coping with high dimensionality and multimodality effectively in Bayesian analysis.
In this paper, we apply DNest to search for continuous GWs in simulated pulsar timing residuals and find that it performs well in terms of accuracy, robustness, and efficiency for a PTA including $\order{\num{e2}}$ pulsars.
DNest also allows a simultaneous search of multiple sources elegantly, which demonstrates its scalability and general applicability. 
Our results show that it is convenient and also high beneficial to include DNest in current toolboxes of PTA analysis.
\end{abstract}

\keywords{gravitational waves; methods: data analysis; pulsars: general}

\section{Introduction}
The first observation of the gravitational wave (GW) signal from a binary black hole merger by LIGO in 2015 has opened up a new window for our exploration of the universe \citep{abbott2016}.
Since then, dozens of GW events from mergers of stellar-mass compact objects have been detected by ground-based interferometers \citep{abbott2019, LIGO2020}, which transforms our understanding of stellar evolution \citep[e.g.][]{marchant2016}, equation of state of extremely dense matter \citep[e.g.][]{abbott2018}, etc.
With ground-based interferometers, we can only observe the high frequency regime ($\sim 10-\SI{e3}{\hertz}$) of the GW universe.
Several missions for space-based interferometers, such as eLISA \citep{lisa2017}, Taiji \citep{hu2017} and TianQin \citep{luo2016}, are in progress to open up the low frequency regime of $\num{e-4}-\SI{e-1}{\hertz}$.
For ultra low frequency GWs ($\sim \num{e-9}-\SI{e-6}{\hertz}$), a natural galactic-scale detector comprised of a network of millisecond pulsars (MSPs), called a pulsar timing array (PTA), is the most promising way \citep{sazhin1978, foster1990, jenet2006}.

The times of arrival (TOAs) of the radio pulses from a rotating MSP can be measured in high precision.
Sophisticated timing models that account for pulsar system dynamics, pulsar-observatory astrometry, dispersion delay and general relativity effects, can be fitted to the observed TOAs of pulses \citep{edwards2006, luo2020}.
Timing residuals are defined as the differences between the observed TOAs and the TOAs predicted by the best-fit model.
The unmodeled effect of a GW passing between the Earth and pulsars is to disturb the background space-time and modulate the propagation time of radio pulses, thus appears coherently across the timing residuals of an array of pulsars, allowing it to be distinguished from noise or other incoherent unmodeled effects.

Searching for the stochastic GW background generated by numerous unresolved close binaries of supermassive black holes (CB-SMBHs) is one of the major scientific objectives of PTAs.
Thanks to the unique quadrupole property of the gravitational radiation, the correlation between timing residuals from a pair of pulsars varies with their angular separation in a distinctive way \citep{hellings1983}.
Recently, the North American Nanohertz Observatory for Gravitational Waves (NANOGrav) has presented the ``12.5yr'' pulsar-timing data set of 47 MSPs \citep{alam2021}.
A common-spectrum stochastic process, described by a power-law, has been found in the timing behaviors across all pulsars, whereas quadrupole spatial correlation of timing residuals between pulsars are still not significant and the smoking gun evidence for the GW background needs further data \citep{arzoumanian202012}.
Besides NANOGrav, major PTA programs in operation includes the European PTA \citep{desvignes2016} and Parkes PTA \citep{kerr2020}, and approximately 100 pulsars have been timed with high precision for GW detection totally.
The number will increase dramatically with the operation of next generation large-scale radio telescope, especially the Five-hundred-meter Aperture Spherical Telescope \citep[FAST;][]{lee2016} and the Square Kilometer Array \citep[SKA;][]{weltman2020}, making a substantial leap in PTA's sensitivity.

Another important goal of PTA programs is searching for signals of continuous GWs from individual CB-SMBHs by fitting the signal model to the timing residuals of pulsars.
Such searches have been conducted in multiple studies using a variety of methods, but no individual sources have been identified yet \citep{zhu2014, babak2016, aggarwal2019}.
It is expected that PTAs including more pulsars timed with higher precision in the future will be sensitive enough to find signals of single resolvable CB-SMBHs in timing residuals.
However, the data analysis with a large number of pulsars will be extremely challenging because of the following reason. 
Since wavelengths of GWs from CB-SMBHs are usually less than uncertainties of distances of pulsars to Earth, each pulsar in the PTA will introduce an unknown pulsar phase parameter in the signal model, making the dimension of the parameter space too high to explore easily.
\cite{ellis2013} introduced a fully Bayesian data analysis pipeline and applied parallel tempering Markov chain Monte Carlo \citep[MCMC, see][]{metropolis1953,hastings1970, swendsen1986, sharma2017} method to sample the posterior probability distribution of model parameters.
NANOGrav applied the method to the observed timing residuals of about $40$ pulsars in their recent studies \citep{aggarwal2019, arzoumanian202009}, but testing with more pulsars, which will certainly be the case in the near future, is needed.
Similarly, \cite{zhu2016} introduced a more efficient frequentist framework and employed particle swarm optimization \citep[PSO, see][]{eberhart1995,wang2010} to search over the parameter space that includes the unknown pulsar phases for the maximum of the likelihood.
Another approach is to reduce the dimension of parameter space and therefore sampling complexity by maximizing \citep{ellis2012,babak2012,wang2015} or marginalizing \citep{taylor2014, wang2017} the likelihood function over some model parameters (especially uninformative pulsar phases) semi-analytically.
It had been used to search for signals of continuous GWs by EPTA \citep{babak2016}, and tested with an extremely large-scale PTA based on the Square Kilometer Array (SKA) era PTA that contains $\num{e3}$ pulsars \citep{wang201704}.
However, the maximization and marginalization rely largely on the analytical form of the signal model, making it not straightforward to generalize to CB-SMBHs with elliptical or evolving orbits without introducing simplified assumptions.
A common assumption is that the pulsar terms do not add up as coherently as Earth terms and so can be discarded for simplicity.
Given that assumption, \cite{babak2012} and \cite{petiteau2013} developed a maximum-likelihood-based with an implementation of the genetic algorithm \citep{holland1975}  to resolve multiple CB-SMBHs.
As a further step, \cite{becsy2020} used trans-dimensional Bayesian inference 
implemented by reversible jump MCMC \citep{green1995} to search for isolate sources and stochastic GW background jointly in PTA data.

To establish a general Bayesian framework for detecting continuous GW with large-scale PTAs, we notice that the diffusive nested sampling (DNest) method proposed by \cite{brewer2011} would be an appropriate choice.
DNest can effectively solve problems arising from high dimensions, multi-modal distributions, highly-correlated parameters and phase changes compared to other sampling methods \citep{brewer2018}.
It has been successfully applied to reconstruct the broad-line region model from reverberation mapping and spectroastrometry data in the research of active galactic nuclei where the dimension of the parameter space can exceed $100$ \citep[e.g.][]{pancoast2014, li2018, wang2020}.

In this work, we develop a code named \texttt{TRAINS}
\footnote{The package name \texttt{TRAINS} is the abbreviation of \emph{Timing Residuals Analysis Integrated with Nested Sampling}.
It can be downloaded via \url{https://github.com/yuyang1995/TRAINS}.}
\citep{songsheng2021} to implement the Bayesian inference with DNest, aiming to model the timing residuals induced by GWs from CB-SMBHs.
We test the capability of DNest on search of nanohertz GW signals generated by a mock population of CB-SMBHs in simulated pulsar timing residuals of a large-scale PTA containing $\num{e2}-\num{e3}$ pulsars.
We fit the signal model to the simulated data and compare probability distributions of model parameters to their input values.
For blind search of a single source, DNest with fully Bayesian framework can overcome the problem of high dimension caused by pulsar phases and perform as well as that with marginalization technique in terms of accuracy, robustness and efficiency.
We further apply the method to search of multiple sources simultaneously and find that it can successfully identify several strongest sources across a wide range of locations and frequencies.
As more and more electromagnetic observational signatures are applied to find CB-SMBH candidates, the targeted search is becoming increasingly important for reliable estimation of orbital parameters by breaking up degeneracies between them \citep{wang2020b}.
As a fully Bayesian method, \texttt{TRAINS} can be easily generalized to analyze pulsar timing residuals and electromagnetic data jointly for targeted searches.

This paper is structured as follows. 
Section 2 introduces the framework used to generate mock data and sample the probability distribution of model parameters.
Section 3 presents searching results in cases of single sources, multiple sources and targeted sources. 
Discussions are provided in section 4, and conclusions are summarized in the last section.

\section{Methodology}
\subsection{Signal model}
The Doppler response to GWs is originally given by \cite{estabrook1975}, which is applied to a binary source firstly in \cite{wahlquist1987}.
We re-derive the timing residuals of a pulsar caused by GWs from a CB-SMBH in Appendix \ref{sec:model} for readers' convenience, but also refer to \cite{aggarwal2019} and references therein.
Here we only present the main results.
The response of a pulsar to a GW source is described by the antenna pattern functions $F^{+}$ and $F^{\times}$,
\begin{equation}
    F^+ = \frac{1}{2} \frac{(\vu*{N}\vdot\vu*{p})^2 - (\vu*{E}\vdot\vu*{p})^2}{1 + \vu*{p}\vdot\vu*{k}} \qc
    F^{\times} = -\frac{(\vu*{E}\vdot\vu*{p})(\vu*{N}\vdot\vu*{p})}{1 + \vu*{p}\vdot\vu*{k}},
\end{equation}
where $\vu*{k}$ is the unit vector pointing from the source to the observer, $\vu*{p}$ is the unit vector pointing from the observer to the pulsar, $\vu*{E}$ and $\vu*{N}$ are two orthogonal normalized basic vectors in the plane perpendicular to $\vu*{k}$, and pointing to the direction of increasing right ascension and declination respectively.

The effect of the GW on pulsar's residuals measured by the observer at time $t$ can be expressed as
\begin{equation}
    s(t) = F^+ \Delta s_+(t) + F^{\times} \Delta s_{\times}(t),
\end{equation}
where $\Delta s_{+,\times}(t) \equiv s_{+,\times}(t) - s_{+,\times}(t_{\rm p})$ is the difference between the so-called Earth term $s_{+,\times}(t)$ and pulsar term $s_{+,\times}(t_{\rm p})$, induced by GW at the Earth and pulsar respectively.
The pulsar time $t_{\rm p}$ is related to the Earth time $t$ as
\begin{equation}
    t_{\rm p} = t - d_{\rm p} (1 + \vu*{p}\vdot\vu*{k})/c \equiv t - \tau_{\rm p},
\end{equation}
where $d_{\rm p}$ is the distance of the pulsar to the observer and $c$ is the speed of light.

For a CB-SMBH in circular orbit, $s_{+,\times}(t)$ is given by
\begin{align}
    s_{+}(t) &= \frac{(G\mathcal{M})^{5/3}}{c^4D_{\rm L}\omega(t)^{1/3}} [\cos2\psi(1+\cos^2\iota)\sin2\varphi(t) - 2\sin2\psi \cos\iota \cos 2\varphi(t)], \nonumber \\
    s_{\times}(t) &= \frac{(G\mathcal{M})^{5/3}}{c^4D_{\rm L}\omega(t)^{1/3}} [\sin2\psi(1+\cos^2\iota)\sin2\varphi(t) + 2\cos2\psi \cos\iota \cos 2\varphi(t)],
\end{align}
where $G$ is gravitational constant, $\mathcal{M}$ and $D_{\rm L}$ is the redshifted chirp mass and luminosity distance of the CB-SMBH, $\iota$ is the inclination angle of the orbital plane, and $\psi$ is the GW polarization angle.
The observed orbital angular frequency $\omega$ of the CB-SMBH evolves slowly as the GW radiates the energy of the binary away gradually,
\begin{equation}\label{eq:orbital_angular_frequency}
    \omega(t) = \omega_0 \left(1 - \frac{t}{t_{\rm merge}}\right)^{-3/8},
\end{equation}
where $\omega_0 \equiv \omega(t = 0)$ is the initial orbital angular frequency and $t_{\rm merge}$ is the merger time,
\begin{equation}
    t_{\rm merge} = \frac{5c^5}{256} (G\mathcal{M})^{-5/3} \omega^{-8/3} = \num{4.4e4} \left(\frac{\mathcal{M}}{\num{e9}M_{\sun}}\right)^{-5/3} \left(\frac{\mathcal{\omega}}{\SI{e-8}{\hertz}}\right)^{-8/3} \si{yr}.
\end{equation}
The variation of the orbital phase with time $\varphi(t)$ is given by
\begin{equation}\label{eq:orbital_phase}
    \varphi(t) = \varphi_0 + \frac{8\omega_0 t_{\rm merge}}{5} \left[1-\left(1-\frac{t}{t_{\rm merge}}\right)^{5/8}\right],
\end{equation}
where $\varphi_0 \equiv \varphi(t = 0)$ is the initial orbital phase.
For the pulsar term evaluated at the pulsar time $t_{\rm p}$, we also define $\omega_{\rm p} \equiv \omega(-\tau)$ and $\varphi_{\rm p} \equiv \varphi(-\tau)$, and we have
\begin{equation}
    \omega(t_{\rm p}) = \omega_{\rm p} \left(1 - \frac{t}{t_{\rm merge} + \tau}\right)^{-3/8} \qc
    \varphi(t_{\rm p}) = \varphi_{\rm p} + \frac{8\omega_{\rm p}(t_{\rm merge} + \tau)}{5} \left[1-\left(1-\frac{t}{t_{\rm merge}+\tau}\right)^{5/8}\right].
\end{equation}
If the time span of the PTA program is much shorter than the merger time $t_{\rm merge}$, orbital angular frequencies in the Earth and pulsar term can be treated as constants respectively, and variations of orbital phases can be approximated as
$\varphi(t) = \varphi_0 + \omega_0 t$ and $\varphi(t_{\rm p}) = \varphi_{\rm p} + \omega_{\rm p} t$ respectively.
If the light travel time between the Earth and pulsar is also much shorter than the merger time, we can further assume $\omega_0 \approx \omega_{\rm p}$ and $\varphi_{\rm p} \approx \varphi_0 - \omega_{0}\tau_{\rm p}$.

\subsection{Simulation of pulsar timing residuals \label{subsec:simu}}
\begin{figure}
    \centering
    \includegraphics[width=\textwidth]{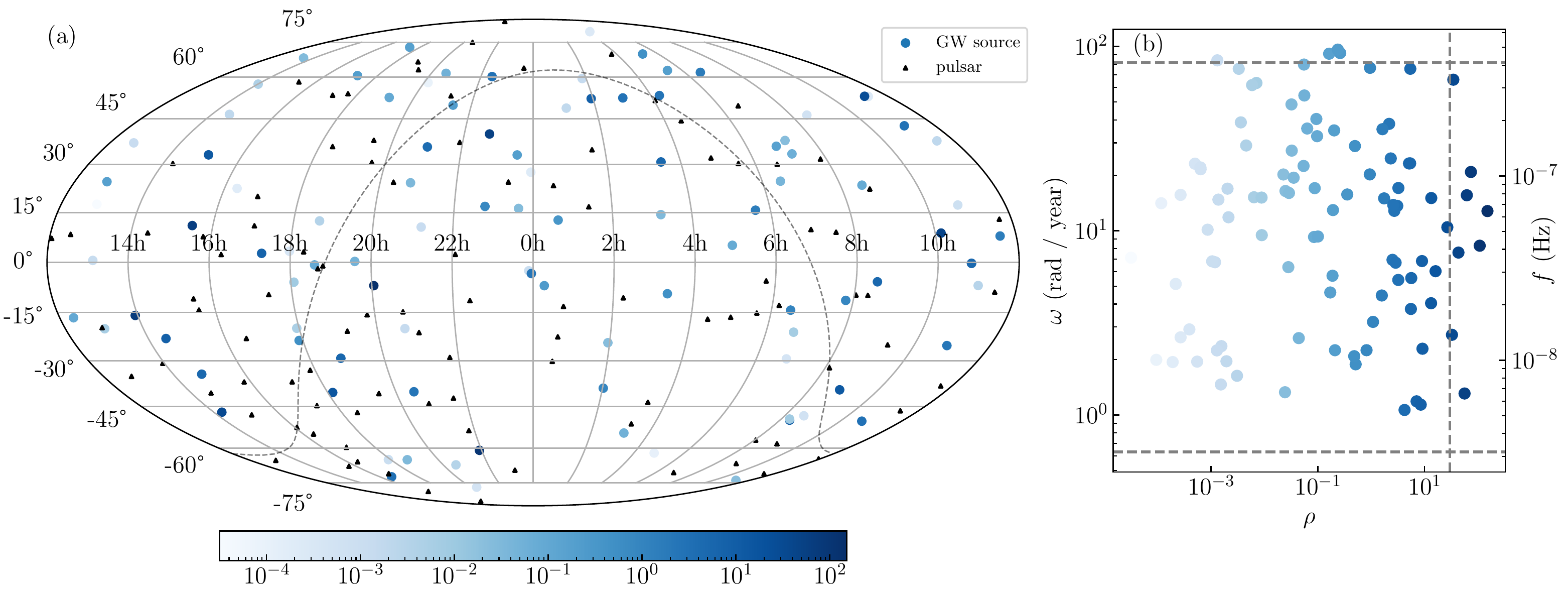}
    \caption{Positions, strength and frequencies of simulated GW sources.
    (a) Distribution of pulsars and CB-SMBHs on the celestial sphere.
    Black triangles trace positions of pulsars while blue dots trace those of CB-SMBHs.
    The depth of blue, as indicated by the colorbar, represents the SNR of the GW signal.
    The dashed gray line denotes the Galactic plane.
    (b) Distribution of SNRs and frequencies of GW sources.
    SNRs and frequencies span from $\num{3e-5}$ to $\num{1.5e2}$ and from $\SI{5e-9}{\Hz}$ to $\SI{5e-7}{\Hz}$ respectively.
    The upper and lower dashed line denotes $f = 1/(2\Delta t)$ (Nyquist frequency) and $f = 1/(2T)$ respectively, where $\delta t$ and $T$ are cadence and duration of the sampling respectively. 
    The vertical dashed line marks $\rho = 30$.
    \label{fig:source}}
\end{figure}

The PTA used for our simulation is composed of $100$ pulsars, provided by the pulsar simulation for the SKA \citep{smits2009}.
Their positions are marked by the black triangles in Fig. \ref{fig:source}(a).
Timing uncertainties of MSPs depends on rotation periods, flux densities and profile widths of pulsars, as well as integration times, band widths and sensitivities of telescopes \citep{wang2018}.
For simplicity, we assume the typical timing uncertainty of each pulsar follows a white Gaussian noise with root mean square value of $\SI{e-7}{\second}$ and it is added to the timing residuals caused by GWs.
Other sources of noise, such as pulsar spin noise, dispersion measure variation and stochastic GW background, have complicated correlations in space or time \citep{tiburzi2016} and could impact CB-SMBH searches in practice \citep{becsy2020}.
As a proof-of-concept of the DNest method, we neglect them here and integrating them into our model will be subjected to a future work.

The GWs are generated by $100$ CB-SMBHs uniformly distributed on the celestial sphere, as shown in blue circles in Fig. \ref{fig:source} (a).
The luminosity distance ($D_{\rm L}$) of the sources range from $\SI{100}{Mpc}$ to $\SI{1000}{Mpc}$.
The cubic of the distance is generated uniformly to achieve a homogeneous distribution of sources in space.
The redshifted chirp mass ($\mathcal{M}$) of the CB-WMBHs follows a log-uniform distribution and its minimum and maximum are $\num{e6}M_{\sun}$ and $\num{e10}M_{\sun}$, respectively.
Observed angular frequencies of GWs ($\omega_{\rm gw}$) are set to range from $\SI{1}{\radian\per\text{yr}}$ to $\SI{100}{\radian\per\text{yr}}$ log-uniformly.
As a result, the characteristic amplitude of timing residuals caused by a single source 
\begin{equation}
\zeta = \frac{(G\mathcal{M})^{5/3}}{c^4 D_{\rm L} (\pi f_{\rm gw})^{1/3}}
= \num{5.5e-8} \left(\frac{\mathcal{M}}{\num{e9}M_{\sun}}\right)^{5/3}
\left(\frac{D_{\rm L}}{\SI{100}{Mpc}}\right)^{-1} 
\left(\frac{f_{\rm gw}}{\SI{e-8}{\hertz}}\right)^{-1/3} \si{\second}
\end{equation}
will span a wide range from $\SI{2.6e-6}{\second}$ to $\SI{1.2e-14}{\second}$.
The cosine of the inclination angle ($\iota$) between the binary orbital plane and the plane of the sky and the GW polarization angle ($\psi$) are uniformly drawn from $[-1,1]$ and $[0, \pi]$, respectively.
Finally, we also assume the initial orbital phase ($\Phi_0$) is uniformly distributed from $0$ to $\pi$.

We emphasis here that our mock sample of CB-SMBHs is quite artificial and does not represent realistic populations in the Universe.
However, this is acceptable since our major goal is to demonstrate the capability of our searching algorithm in large-scale PTAs rather than to make robust predictions for realistic observations.
We also neglect evolution of binary orbits in order to compare the method with the technique of maximization and marginalization, but a test of the method in the regime of evolving orbits will be presented in Appendix \ref{sec:evolution}.

Given the information of pulsars and CB-SMBHs, timing residuals for each pulsar can be calculated by the signal model of non-evolving binaries.
The timing residual data are sampled with cadence of $14$ days for a period of $5$ years.
The GW angular frequencies of the most sources are below the Nyquist frequency of the sampling to avoid frequency leak, as shown in Fig. \ref{fig:source}(b).
However, frequency reach of PTA-based GW search can be extended far beyond the Nyquist frequency associated  with the single pulsar's cadence by exploiting asynchronous observations from multiple pulsars \citep{wang2021}, which will be explored in the future.

To quantify the relative strength of the signal for each source, we define the network signal-to-noise ratio (SNR) $\rho$ for a source $i$ as
\begin{equation}
    \rho_i = \left[\sum_{I=0}^{N_{\rm p}} \sum_{k=0}^{N_{\rm t}} \left(\frac{s_{iI}(t_k)}{\sigma_I}\right)^2\right]^{1/2},
\end{equation}
where $s_{iI}(t_k)$ is the timing residual generated by source $i$ for pulsar $I$ at time $t_k$, $\sigma_I$ is the timing uncertainty of pulsar $I$, and $N_{\rm p}$ and $N_{\rm t}$ are number of pulsars and observation times respectively. 
In our simulated data, SNRs of CB-SMBHs span from $\num{1.3e-5}$ to $\num{1.5e2}$, as shown in Fig. \ref{fig:source}(b), and $8$ sources have SNRs above $30$.

\subsection{Bayesian analysis}
In order to fit the non-evolving CB-SMBH model with $N_{\rm s}$ sources to the data, we have $7$ free parameters for each source, as summarized in Table \ref{tab:CB-SMBH}.
Note that the chirp mass $\mathcal{M}$ and luminosity distance $D_{\rm L}$ of the source are degenerate and only the amplitude of timing residuals $\zeta$ is included in the fitting.
When calculating timing residuals for each pulsar, we need to figure out a pulsar phase for each pair of source and pulsar,
\begin{equation}
    \varphi_{\rm p} = \varphi_0 - \frac{\omega_{\rm gw} d_{\rm p} (1 - \cos\theta)}{2c}, 
\end{equation}
where $\theta$ is the opening angle between the source and the pulsar subtended at the observer.
However, uncertainties of $d_{\rm p}$ are much larger than the GW wavelengths when dealing with real data.
We therefore treat each pulsar phase as a free parameter with a uniform prior in our mock data analysis \citep{corbin2010}.

Now, we have $N_{\rm s} (7 + N_{\rm p})$ free parameters in our models.
The likelihood function of data set $\mathscr{D}$ for parameter $\{\bm{\Theta}\}$ in model $M$ is therefore
\begin{equation}\label{eq:likelihood}
    P(\mathscr{D}|\bm{\Theta},M) = \prod_{I=1}^{N_{\rm p}} \prod_{k=1}^{N_{\rm t}} \frac{1}{\sqrt{2\pi \sigma^2}} \exp{-\frac{[r_I(t_k) - r_{I,M}(t_k,\bm{\Theta})]^2}{2\sigma_I^2}},
\end{equation}
where $N_t$ is the number of data points in timing residuals for each pulsar, $\sigma_I$ is the timing uncertainty of the pulsar, and $r_I(t_k)$ and $r_{I,M}(t_k,\bm{\Theta})$ are timing residual for pulsar $I$ at timing $t_k$ of  data and model, respectively.
In light of Bayes' theorem, the posterior probability distribution for $\bm{\Theta}$ is given by
\begin{equation}\label{eq:posterior}
    P(\bm{\Theta} | \mathscr{D},M) = \frac{P(\bm{\Theta}|M) P(\mathscr{D} | \bm{\Theta},M)}{P(\mathscr{D}|M)},
\end{equation}
where $P(\bm{\Theta}|M)$ is the prior distribution of the model parameter and $P(\mathscr{D}|M)$ is a normalization factor.

For $N_{\rm s} = 1$, if the seven parameters of the source $\bm{s}$ are given, the likelihood function Eq. \ref{eq:likelihood} can be maximized analytically over pulsar phase parameters $\bm{\phi}$ \citep{wang2015}. If we use the new likelihood function
\begin{equation}
    P(\mathscr{D}|\bm{s},M) = \max_{\{\bm{\phi}\}}P(\mathscr{D}|\bm{\Theta},M) \qc \bm{\Theta} = \{\bm{s}, \bm{\phi}\}
\end{equation}
instead of Eq. \ref{eq:likelihood}, the dimension of the parameter space will be reduced from $N_{\rm p} + 7$ to $7$.
In such a case, traditional optimization or sampling algorithm, such as PSO and MCMC, can be applied to find the parameter $\bm{s}$ of the source.

A more mathematically rigorous way to get the probability distribution of a subset $\bm{s}$ of model parameters $\bm{\Theta}$ is to marginalize nuisance parameters $\bm{\phi}$ out, namely,
\begin{equation}\label{eq:marginalization}
    P(\bm{s} | \mathscr{D},M) = \int P(\bm{\Theta} | \mathscr{D},M) \dd{\bm{\phi}} = \frac{P(\bm{\bm{s}}|M)}{P(\mathscr{D}|M)} \int P(\mathscr{D} | \bm{\Theta},M) P(\bm{\bm{\phi}}|M) \dd{\bm{\phi}}.
\end{equation}
Here, we assume the prior of source parameters $\bm{s}$ and that of pulsar phase parameters $\bm{\phi}$ are independent.
When $N_{\rm s} = 1$, the integration at the right hand side of Eq. (\ref{eq:marginalization}) can be worked out semi-analytically and integrated numerically \citep{wang2017}.
Again, the dimension of the parameter space reduces to $7$.

However, if $N_{\rm s} > 1$, or the orbit of the CB-SMBH is elliptical or evolving, the maximization or marginalization of likelihood function over pulsar phases cannot be performed analytically. 
The dimension of the parameter space will be very large ($\order{10^2} - \order{10^3}$) as the number of available pulsars in timing array increases.
In this case, the DNest  method \citep{brewer2011}
\footnote{
    Original implementation of the algorithm developed by \cite{brewer2011} is available at \url{https://github.com/eggplantbren/DNest4}. 
    In this work, we use our own DNest package \texttt{CDNest} \citep{li2020} that is written in C language and enables the standardized parallel message passing interface, which is available at \url{https://github.com/LiyrAstroph/CDNest}.}
will be an appropriate choice to tackle the challenge of high dimensionality.

\begin{deluxetable}{llll}
    \tablecaption{Parameters used in the CB-SMBH model \label{tab:CB-SMBH}}
    \tablewidth{0pt}
    \tablehead{\colhead{Parameters} & \colhead{Meanings} & \colhead{Prior ranges} & \colhead{Prior probability}}
    \startdata
    $\alpha$ & right ascension of the CB-SMBH & $[0, 2\pi]$ & uniform \\
    $\delta$ & declination of the CB-SMBH & $[-\pi/2,\pi/2]$ & $\sin\delta$ uniform \\
    $\iota$ & inclination of the orbital plane & $[0, \pi]$ & $\cos\iota$ uniform \\
    $\psi$ & GW polarization angle & $[0, \pi]$ & uniform \\
    $\varphi_0$ & initial orbital phase & $[0, \pi]$ & uniform \\
    $\zeta$ & amplitude of timing residuals & $[\num{e-15},\num{e-6}]\si{\second}$ & log uniform \\
    $\omega_{\rm gw}$ & observed angular frequency of GW & $[1,100]\si{\radian.yr^{-1}} $ & log uniform \\
    $\varphi_{\rm p}$ & pulsar phase & $[0, \pi]$ & uniform 
    \enddata
\end{deluxetable}

\subsection{Diffusive nested sampling}
To evaluate the evidence $Z$ of a model $M$ for data set $\mathscr{D}$:
\begin{equation}\label{eq:evidence}
    Z \equiv P(\mathscr{D}|M) = \int P(\mathscr{D} | \bm{\Theta},M) P(\bm{\Theta}|M) \dd{\bm{\Theta}},
\end{equation}
\cite{skilling2004} proposed the nested sampling method.
Note that
\begin{equation}
    P(M|\mathscr{D}) = \frac{P(\mathscr{D}|M)P(M)}{P(\mathscr{D})}.
\end{equation}
It is generally reasonable to assign equal priors for models under consideration, as a result, the evidence can be directly used for model selection. 

Nested sampling firstly samples $n$ particles $\bm{\Theta}_i$ from the prior $P(\bm{\Theta}|M)$ and evaluates the likelihood $L(\bm{\Theta}_i) \equiv P(\mathscr{D} | \bm{\Theta}_i,M)$ of each point. 
The particle with the lowest likelihood $L_1$ is recorded and replaced by a new one drawn from the prior but under a constraint $L(\bm{\Theta}) > L_1$ via MCMC.
Mathematically, the new prior can be expressed as
\begin{equation}
    p_j(\bm{\Theta}) = \frac{P(\bm{\Theta}|M)}{X_j}
    \begin{cases}
        1 \qif P(\mathscr{D} | \bm{\Theta},M) > L_j \\
        0 \qif P(\mathscr{D} | \bm{\Theta},M) <= L_j
    \end{cases}
    \qc
    X_j \equiv \int_{L(\bm{\Theta}) > L_j} P(\bm{\Theta}|M) \dd{\bm{\Theta}}
\end{equation}
(here $j=1$).
Then we record the minimum of likelihoods $L_2$ and repeat the process to build up nested likelihood levels $L_1 < L_2 < \cdots$.
It can be proved that prior probability enclosed by $j$th level $X_j$ has an expectation of $E(X_j) = \exp(-j/n)$.
Consequently, the evidence can be approximated by
\begin{equation}
    Z \approx \sum_{j} L_j(X_{j-1}-X_{j}). 
\end{equation}
As particles are constrained to higher levels with higher likelihoods, a posterior sample of $\bm{\Theta}$ can also been obtained from those recorded points with the lowest likelihoods in each iteration \citep{skilling2004}.

However, classical nested sampling performs unsatisfactorily when sampling multi-modal or highly correlated distributions.
Particles may be stuck in local maximum and fail to explore the whole parameter space if a distribution has isolate ``islands'' with high likelihoods.
DNest makes improvements by assigning a label $j$ to each particle indicating which particular level it is currently constrained by, where $j=0,1,\cdots,j_{\rm max}$ and $j_{\rm max}$ is the current top level.
Instead of sampling all points according to the restricted prior $p_{j_{\rm max}}$, DNest samples particles' positions $\bm{\Theta}$ and labels $j$ concurrently by a mixed distribution
\begin{equation}
    p(\bm{\Theta},j) = w_j p_j(\theta),
\end{equation}
where $w_j$ is the chosen weighting scheme.
In the process of creating levels, the exponentially-decaying weights
\begin{equation}
    w_j \propto \exp(\frac{j-j_{\rm max}}{\Lambda})
\end{equation}
are adopted, where $\Lambda$ is a backtrack length controlling how far particles can diffuse to lower levels.
As a result, particles can diffuse to lower levels where the parameter space is more connected so that particles can easily explore the whole space.
The weights are adjusted to uniform once the desired number of levels has been generated, to further sample the posterior distribution of $\bm{\Theta}$.

If levels created are insufficient, we may not see the peak in the series $L_j(X_{j-1} - X_{j})$, i.e., we have not found the level that contains the most posterior probability to obtain a large enough posterior sample.
In this case, we can introduce a ``temperature'' $T$ and modify the likelihood of each level to $\tilde{L}_j = L_j \exp(-T)$.
If the peak appears in the series $\tilde{L}_j(X_{j-1} - X_{j})$, we can get a posterior sample of the modified posterior probability, which is a broadened version of the original one.
The uncertainties of model parameters would be overestimated, but it can still extract the information of parameters from data as much as possible even though the standard sampling with $T=1$ fails to find out the spiky peak of the posterior probability.

\section{Results}
\subsection{Blind search: a single source \label{subsec:blind_single}}

\begin{figure}
    \plotone{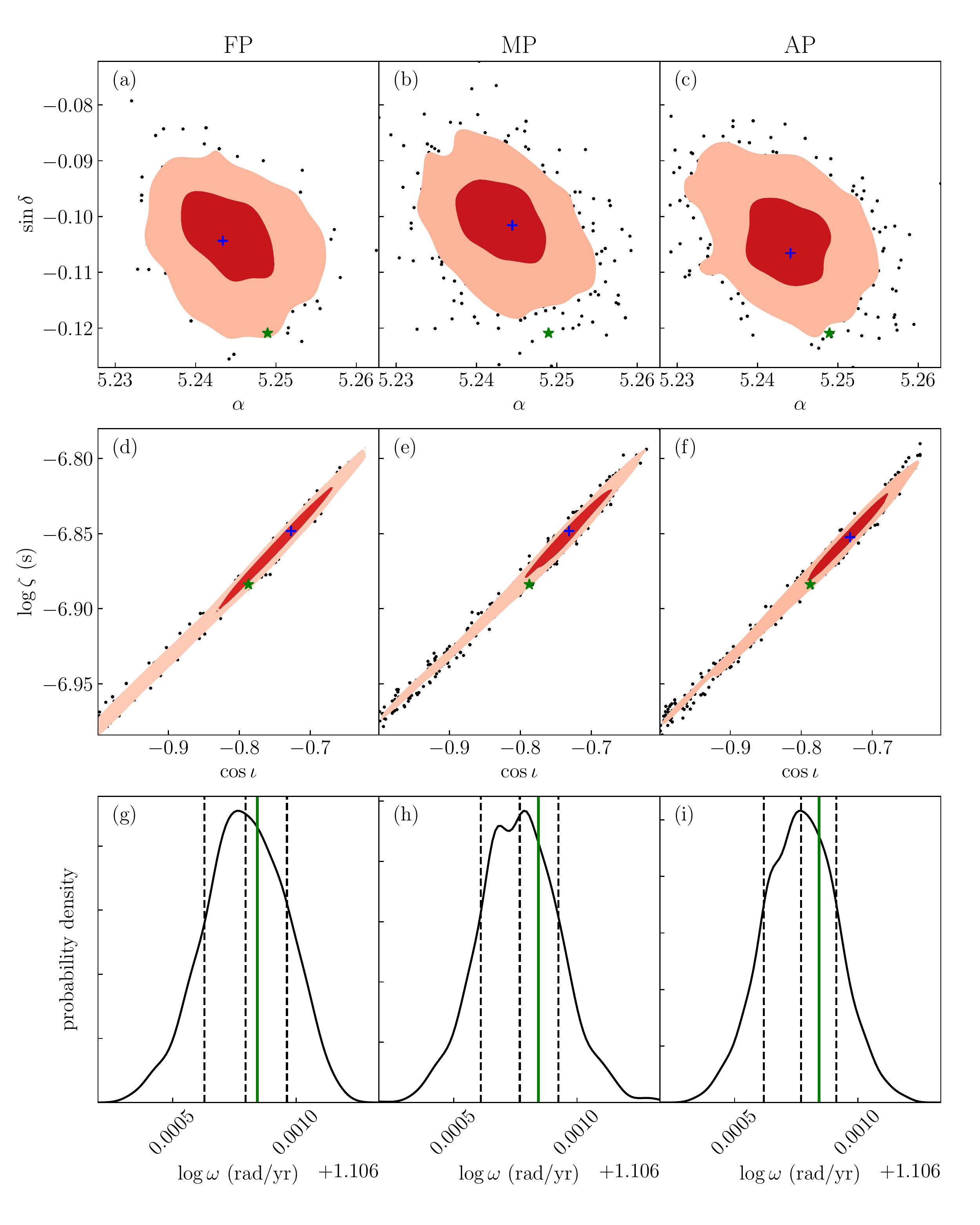}
    \caption{Posterior probability distributions of some model parameters of the GW source. 
    The first, second and third columns are distributions reconstructed by FP, MP and AP methods respectively.
    The first row shows distributions of the right accession and declination of the source. 
    The red and pink colors represent the $1\sigma$ and $2\sigma$ regions of the distribution.
    Black dots are posterior samples outside the $2\sigma$ region.
    The green star marks the input parameters, while the blue cross indicates the peak of the distribution.
    The second row illustrates distributions of the inclination and amplitude of the source.
    The last row shows probability density functions of the angular frequency of the source.
    The dashed lines are the 16\%, 50\% and 84\% quantiles, while the green line is the input frequency.
    All probability density functions are calculated by \texttt{gaussian\_kde} method of \texttt{scipy.stats} module \citep{virtanen2020} of \texttt{Python} from posterior samples reconstructed by DNest method.
    \label{fig:single}}
\end{figure}

Firstly, we fit the CB-SMBH model containing one source to the pulsar timing residuals using three different methods: 
(1) treating pulsar phases as free parameters (FP); 
(2) maximizing the likelihood function over pulsar phases (MP); 
(3) averaging the likelihood function over pulsar phases (AP).
The posterior probability distributions in the three cases are all sampled by DNest algorithm.
The CB-SMBH with the highest SNR ($\rho = 151$) is identified as we expected.
Posterior distributions of major parameters of the source are illustrated in Fig. \ref{fig:single}.
Uncertainties and biases of the recovered source's position are also listed in the first row of Table \ref{tab:bsss}.

\begin{deluxetable}{CCCCCCC}
    \tablecaption{Uncertainties and biases of the recovered source's position for different SNRs and searching methods \label{tab:bsss}}
    \tablewidth{0pt}
    \tablehead{
    \colhead{SNR} & \multicolumn{3}{c}{Uncertainty ($\si{deg^2}$)} & \multicolumn{3}{c}{Bias ($\si{deg}$)} \\
    \cline{2-7}
    \colhead{} & \colhead{FP} & \colhead{MP} & \colhead{AP} & \colhead{FP} & \colhead{MP} & \colhead{AP}
    }
    \startdata
    151 & 1.75 & 1.61 & 1.78 & 1.00 & 1.14 & 0.87 \\
    30 & 52.9 & 65.5 & 49.4 & 3.25 & 10.4 & 2.14 \\
    10 & 223 & 11026 & 233 & 6.07 & 130 & 5.17
    \enddata
    \tablecomments{The uncertainty of the source's position is definded as the size of the $2\sigma$ region of the posterior sample (enclosing $86\%$ of the total probability), and the bias is defined as the angular distance between the input location and the center of the posterior distribution.}
\end{deluxetable}

The first row of Fig. \ref{fig:single} shows the distribution of right ascension and declination of the source.
For FP and AP methods, the actual location of the source lies near or within the $2\sigma$ region (enclosing $86\%$ of the total probability) of the posterior sample, while the deviation in MP method is a little bit larger.
The second row shows the distribution of the amplitude of timing residuals and the inclination of the orbital plane.
The input value is at or near the edge of the $1\sigma$ contour (enclosing $40\%$ of the total probability), and the long tail of the distribution might be caused by the strong correlation between inclinations and amplitudes.
The last row displays the probability density of the angular frequency of the GW.
The input frequencies lies within the $1\sigma$ range of the distribution, and the uncertainty is only $\SI{2e-4}{dex}$, making it the best determined model parameters.
In a word, DNest can effectively cope with high dimensionality owing to pulsar phases when the SNR of the GW source is relatively high.

\begin{figure}
    \plotone{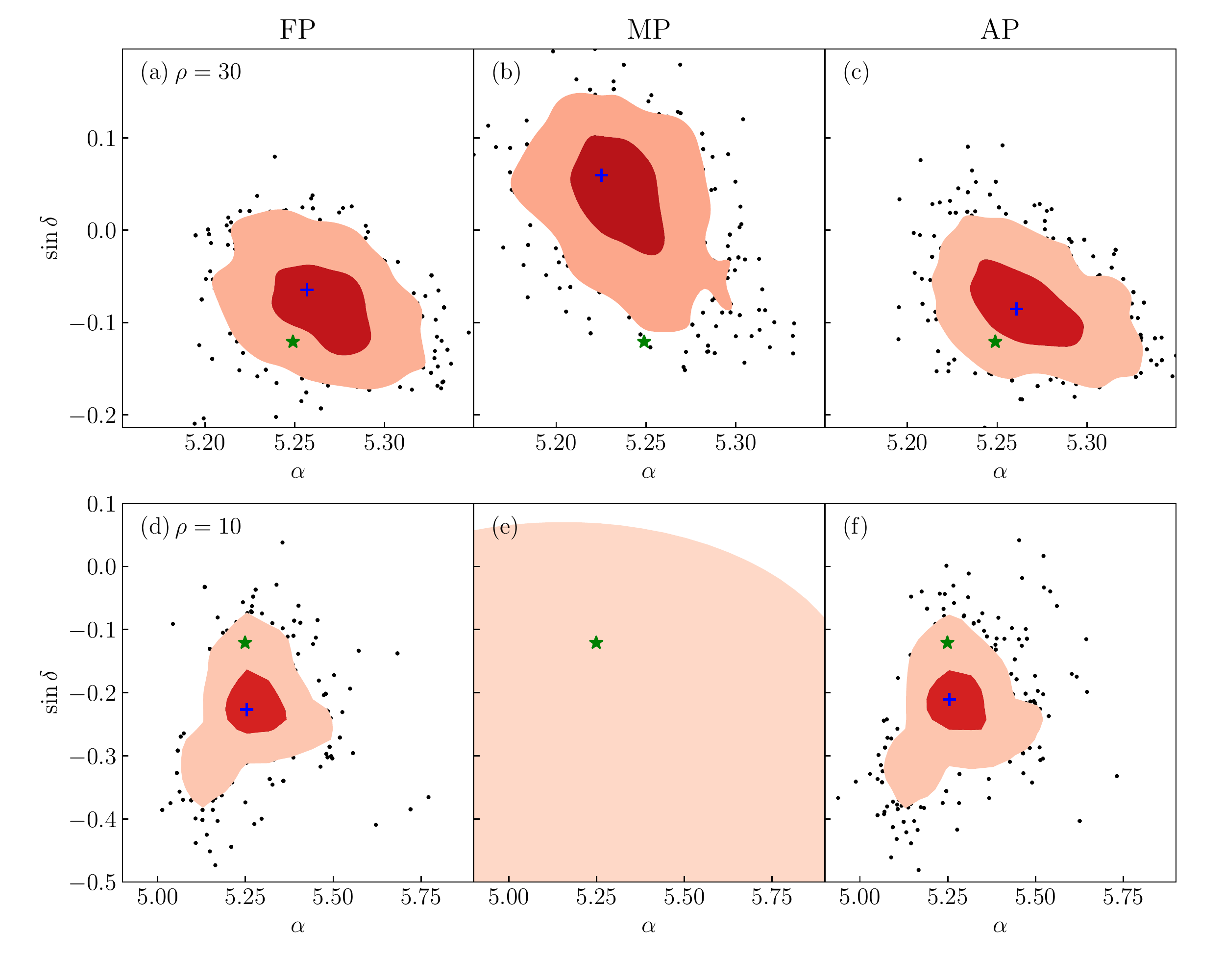}
    \caption{Posterior probability distributions of the right ascension and declination of the GW source.
    The first, second and third column are distributions reconstructed by FP, MP and AP methods respectively.
    The red and pink colors represent the $1\sigma$ and $2\sigma$ regions of the distribution.
    Black dots are posterior samples outside the $2\sigma$ region.
    The green star marks the input parameters, while the blue cross indicates the peak of the distribution.
    In the first row, the SNR of the source is $30$, while in the second row, the SNR is only $10$.
    Note that in sub-figure (e), the distribution is very broad and only a portion is shown.
    \label{fig:weak}}
\end{figure}

In order to test the performance of DNest in search of relatively weak source, we increase the timing uncertainties of all pulsars to $\SI{5e-7}{\second}$, bringing down the SNR of the strongest source to $30$.
As can be seen from the first row of Fig. \ref{fig:weak} and the second row of Table \ref{tab:bsss}, the source can still be located, but with larger uncertainties and biases.
Posterior distributions of CB-SMBH's location obtained by FP and AP methods are similar.
Distribution obtained by MP method is a little deformed and has a slightly larger uncertainty and bias.
We further increase the timing uncertainties to $\SI{1.5e-6}{\second}$, reducing the SNR to $10$.
The result is illustrated in the second row of Fig. \ref{fig:weak} and the third row of Table \ref{tab:bsss}.
Both FP and AP methods work well, but the posterior distribution recovered by MP method is extremely scattered, failing to locate the source.
The failure of MP in the case of low SNR may be related to ill-posedness of the inverse problem of estimating the large number of pulsar phase parameters \citep[e.g.][]{wang2015}.
The likelihood function is expected to be highly degenerate over the pulsar phases and contain strong secondary maxima.
As SNR decreases, the locations of such secondary maxima are more likely to become the global maximum under small perturbations from the noise in the data.
The jumping of locations of global maximum to radically different values will lead to a large bias and uncertainty in parameter estimation.
We conclude that the performance of DNest with FP in high dimensional parameter space is still competitive even if the signal of the source is weak.

In the process of getting the results above, the FP and MP methods are the most efficient, but the MP method performs poorly in the case of weak signals.
The AP method takes about three times as long as the FP and MP methods.
The FP method with DNest is not only efficient and robust, but also applicable to a wider range of cases without approximations, such as elliptical orbits, slowly evolving orbits, especially the simultaneous search of multiple sources. 
We emphasize here that MP and AP are originally derived from the maximum likelihood approach, which produces point estimates of parameters rather than samples from their posterior distributions. 
Maximization can be performed with fast heuristic optimizers, making point estimation with MP and AP much more efficient.

\subsection{Blind search: multiple sources}
In this section, we try to locate all CB-SMBHs with $\rho > 30$ simultaneously from pulsar timing residuals.
To this end, we include eight GW sources in our model, each with its own pulsar phases, increasing the dimension of the parameter space to $856$.
We note that if we switch the order of sources in the model, the signal remains unchanged.
To avoid this ambiguity, we demand that $\omega_1 < \cdots < \omega_i < \cdots < \omega_8$, where $\omega_i$ is the angular frequency of GW generated by the $i$th source.
This can be achieved by sorting angular frequencies of all sources after updating particles' positions in parameter space via MCMC.

\begin{figure}
    \centering
    \includegraphics[width=\textwidth]{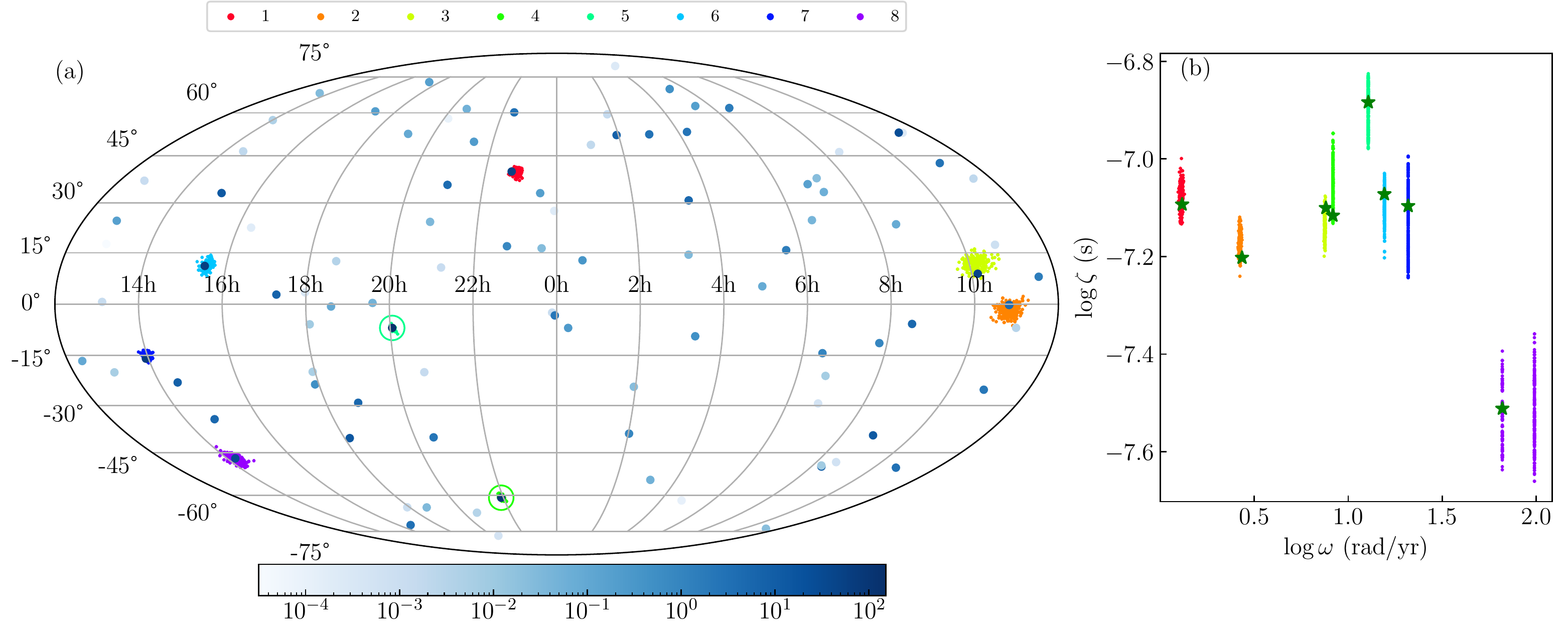}
    \caption{Posterior samples of some model parameters of all GW sources found by DNest.
    (a) Posterior samples of locations of different GW sources on celestial sphere. 
    Large blue dots mark positions of CB-SMBHs.
    The depth of blue, as indicated by the colorbar, represents the SNR of the GW signal.
    Small colorful dots mark posterior samples of right accessions and declinations of different sources.
    Eight sources with highest SNRs are identified.
    The posterior distributions of two sources are too concentrated to recognize in the figure. We draw two circles around them to indicate their positions.
    (b) Posterior samples of amplitudes and angular frequencies of different GW sources.
    Green stars mark true values while colorful dots mark posterior samples.
    Different sources are sorted by their frequencies.
    \label{fig:multi}}
\end{figure}

Posterior samples of model parameters of all GW sources found by DNest are shown in Fig. \ref{fig:multi}.
As illustrated in Fig. \ref{fig:multi}(a), eight sources with highest SNRs are identified.
The higher the SNR, the more accurate the location of the GW source.
Fig. \ref{fig:multi}(b) presents posterior samples of amplitudes and angular frequencies of these sources.
Angular frequencies of different sources are well separated in posterior samples, implying that it is appropriate to eliminate ambiguity of exchanging sources by sorting frequencies.
There are double peaks in the posterior sample of the frequency of the $8$th source, and the minor peak matches the input frequency.
The frequency of the $8$th source is close to the Nyquist frequency of the data sampling, raising the problem of aliasing.
We also notice that the frequency uncertainties of the first two sources are relatively large when compared to those of other sources, probably because the signals they produce change too slowly to have enough cycles for precise measurement of periods.

\begin{figure}
    \centering
    \includegraphics[width=\textwidth]{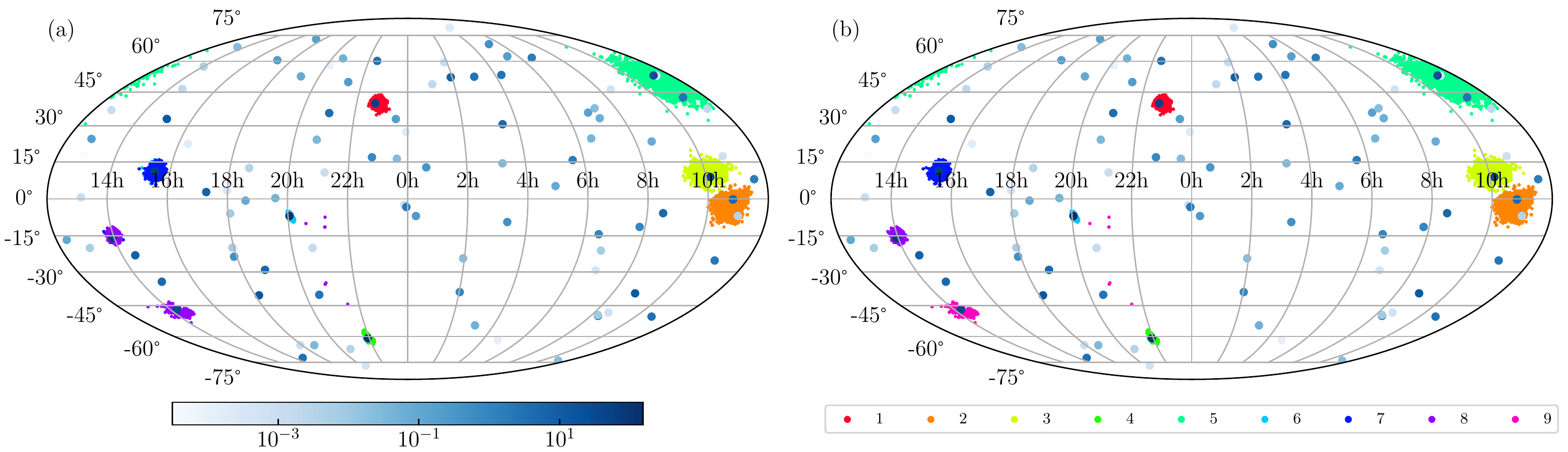}
    \caption{Posterior samples of locations of all GW sources found by DNest.
    Large blue dots mark positions of CB-SMBHs. 
    Small colorful dots mark posterior samples of right accessions and declinations of different sources.
    (a) Before clustering, colors of small dots only represents labels of the parameters in the model, which may not correspond to different sources one to one, i.e., one color may correspond to multiple sources, or different colors may correspond to one source.
    (b) After clustering, colors of small dots correspond to different sources one to one.
    The clustering is performed by \texttt{KMeans} method of the \texttt{Python} module  \texttt{sklearn.cluster} \citep{pedregosa2011}.
    \label{fig:multi2}}
\end{figure}

An interesting fact happens if we ``flatten'' the posterior probability distribution artificially by setting the ``temperature'' of DNest to $T=2.3$.
The posterior probability density reaches its maximum when parameters of the eight sources in our model match those of the eight strongest sources.
However, the probability also has a local maximum if parameters of the eighth strongest source are replaced by those of the ninth strongest one.
This minor peak can be sampled if we ``flatten'' the posterior probability distribution properly.
In the posterior samples, some points correspond to the major peak while others to the minor peak.
Therefore, parameters of a particular source may have different positions in these two kinds of parameter vectors.
For example, as shown in Fig. \ref{fig:multi2}(a), the source at $\sim (\SI{16}{h}, \SI{15}{\degree})$ are sampled by both blue and cyan dots.

Sources in the posterior sample can be clustered by K-means algorithm \citep{lloyd1982} according to their right ascensions, declinations and frequencies.
The result is shown in Fig. \ref{fig:multi2}(b).
Evidently, nine sources are identified, including the source at the right upper corner which have not been identified in Fig. \ref{fig:multi}.
The SNR of the newly identified source is $26$.
By increasing the ``temperature'' of DNest properly, we can even find more sources than there are in the model, but at a cost of larger uncertainties.

Generally speaking, DNest shows its great power in multi-source search, despite the huge dimension of the parameter space.

\subsection{Targeted search}
In blind search, the strongest source are always found first.
To search for a relatively weak source in timing residuals, we have three methods:
(1) Search for the weak source with strong sources simultaneously;
(2) Search for strong sources first, then subtract their signals and search for the weak source;
(3) Use the electromagnetic information to conduct targeted search for known CB-SMBH candidates.
The first two methods spend most of their time in resolving strong sources. If we are only interested in a known candidates, the last method will be the most efficient one.

Currently, there are more than 100 CB-SMBH candidates identified by characteristic electromagnetic information, such as periodic light curves \citep[e.g.][]{graham2015, charisi2016} and emission line profiles \citep[e.g.][]{bon2012,eracleous2012,li2016,li2019}.
Detectability of CB-SMBHs traced by periodic light curves with realistic PTAs has also been studied \citep{xin2021}.
Unlike blind search, the location of the source to be identified is already known from electromagnetic observations.
As an example, we try to search for the source at $(\SI{11}{h}, \SI{0}{\degree})$, i.e., the source identified by orange dots in Fig. \ref{fig:multi}(a).
The SNR of the source is about $32$.

\begin{figure}
    \centering
    \includegraphics[width=\textwidth]{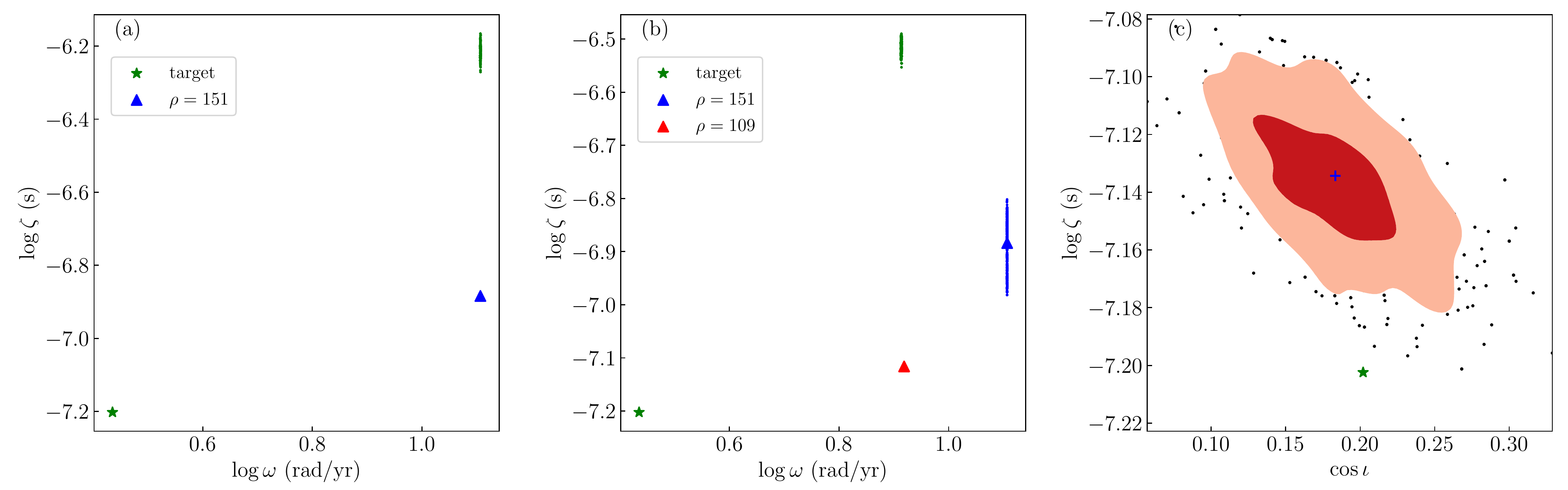}
    \caption{Posterior samples of part of parameters in targeted search.
    (a) Posterior samples obtained with fixed right ascension and declination in the model.
    Green dots are posterior samples of the amplitude and angular frequency of the target while the green star marks the true values.
    The blue triangle indicates the amplitude and frequency of the source with the highest SNR.
    (b) Posterior samples obtained by including one more source in the model.
    Green and blue dots are posterior samples of the target and the strongest source respectively.
    The red triangle marks the second strongest source.
    (c)  Posterior samples obtained with fixed right ascension, declination and angular frequency in the model.
    The red and pink colors represent the $1\sigma$ and $2\sigma$ regions of the distribution of the inclination and amplitude.
    Black dots are posterior samples outside the $2\sigma$ region.
    The green star marks the input values, while the blue cross indicates the peak of the distribution.
    \label{fig:target}}
\end{figure}

Firstly, we fix the right ascension and declination of the source in the model, and fit the model to the timing residuals using DNest method.
The result is shown in Fig. \ref{fig:target}(a).
The posterior sample of the amplitude and frequency of the source does not match those of the target at all.
The reconstructed amplitude is an order of magnitude larger than the true value, and the reconstructed frequency is close to that of the source with the highest SNR.
In the process of fitting, the model attempts to match the overall amplitude and main periodic characteristics of timing residuals, though with misplaced phases.
The difference of this ``fake'' signal to the data is still much smaller than the difference of the true signal with a small amplitude to the data.
To avoid the impact of the strongest source on the targeted search, we include one more source in the signal model.
The result is shown in Fig. \ref{fig:target}(b).
The strongest source has been successfully resolved.
However, the reconstructed amplitude of the target is still an order of magnitude larger, and the recovered frequency matches that of the second strongest source.
We can infer that the target can be resolved only if all stronger sources are included in the signal model.

Now we further fix the frequency of the source in the model, since the orbital period of the CB-SMBH candidate can be constrained by periodic signals in electromagnetic observations.
The result is shown in Fig. \ref{fig:target}(c).
The true amplitude and inclination of the target lies outside the $2\sigma$ region of the posterior sample, but the difference between the true amplitude and the peak value of the posterior distribution is still within $\SI{0.1}{dex}$.

The location of the CB-SMBH candidate can always be determined precisely from electromagnetic observations.
However, the orbital period of the candidate can not be determined directly if it is identified by non-periodic characteristics such as velocity-delay map from reverberation mapping \citep[e.g.][]{wangjm2018} or differential phase curves from spectroastrometry \citep[e.g.][]{songsheng2019}.
Consequently, we should fit the CB-SMBH model to time residuals and electromagnetic data jointly to search for the GW signal.
DNest is also suitable for such joint analysis \citep[e.g.][]{wang2020}.

\section{Discussions}
\subsection{Determining the number of GW sources}
A major concern when searching for multiple sources simultaneously is to determine the number of GW sources in the model.
It can be considered as a problem of model selection in the Bayesian framework.
Fortunately, as a variant of the nested sampling method, DNest can evaluate the evidence of the model directly (see Eq. \ref{eq:evidence}).
Therefore, we can increase the number of GW sources in the model gradually and sampling the probability distribution of model parameters with DNest until the evidence reaches its peak value.
In such iterations, the information obtained in the present runs can be used to the next run.
To be specific, when adding a new source to the model, the ranges of parameters of previously found sources can be further restricted according to their sampled posteriors.
In practice, we can set the upper (lower) bound of the parameter to be its mean plus (minus) five times the standard deviation, and the likelihood beyond this range can be neglected.
We note that the new prior density of the parameter will be $L/L^{\prime}$ times higher, where $L$ and $L^{\prime}$ is the original and modified span of the parameter's value respectively.
As a result, the actual evidence of the model will be $V/V^{\prime}$ times lower, where $V$ and $V^{\prime}$ is the original and modified prior volume respectively.
In this way, the information obtained with the previous model is being utilized, accelerating the sampling significantly.
However, if there are GW sources with very close SNR, the posterior sample of the newly added source in the model may have multiple peaks, corresponding to different sources.
Clustering algorithms must be applied to separate them in order to get real means and standard deviations of sources' parameters.

\begin{figure}
    \centering
    \includegraphics[width=\textwidth]{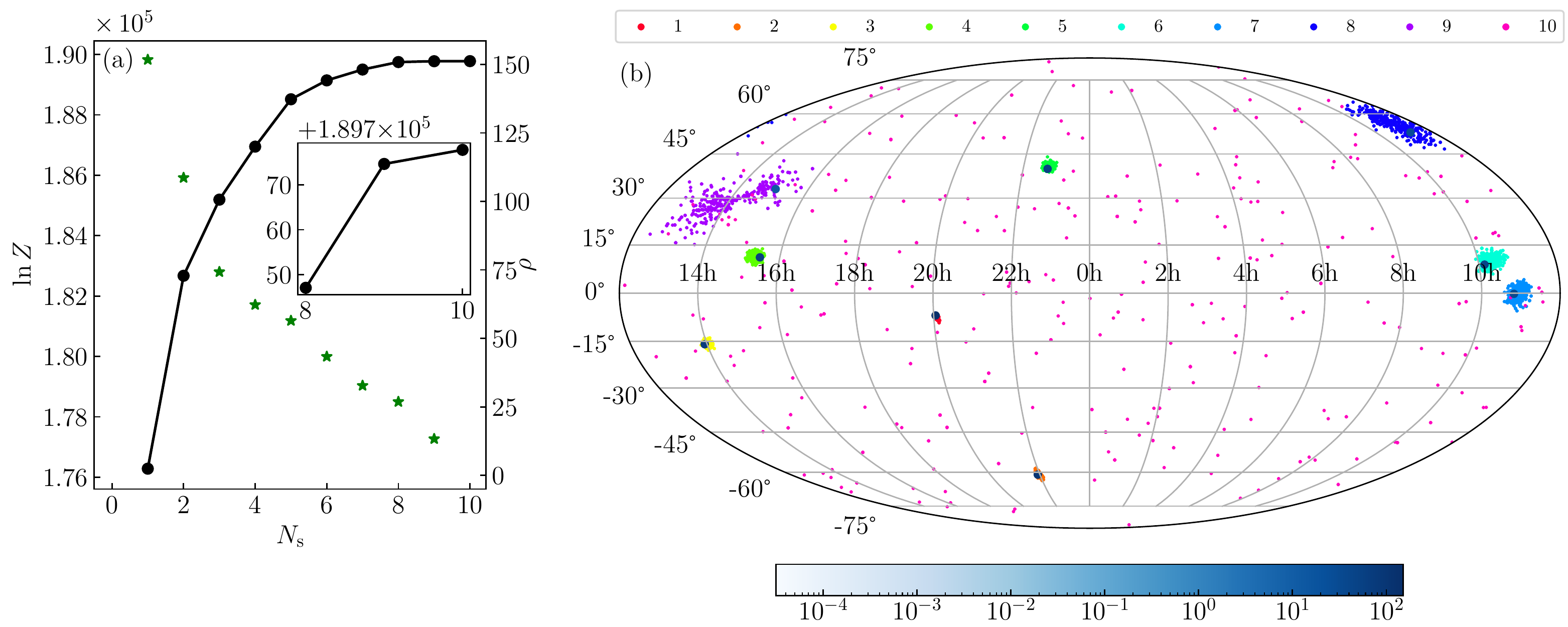}
    \caption{(a) Logarithm of evidences of signal models with different number of GW sources.
    The black line describes the variation of Bayesian evidence as the number of GW sources in the signal model increases.
    Green stars mark SNRs of ten sources used to generate the timing residuals.
    (b) Posterior samples of locations of all GW sources found by DNest.
    Large blue dots and small colorful dots mark positions of input CB-SMBHs and their posterior distributions respectively.
    \label{fig:evidence}}
\end{figure}

As a preliminary test, we avoid this complexity by selecting $9$ CB-SMBHs with quite different SNRs from our mock sample to generate pulsar timing residuals.
Then evidences for signal models with different number of sources can be evaluated as we described above.
The result is shown in Fig. \ref{fig:evidence}.
The model evidence grows gradually as the number of sources in the model increases, but the increase rate drops with the decrease of the SNR of the newly added source.
We note that the evidence ratio between the $10$-source and $9$-source model is about $23$, which seems to supports the $10$-source model strongly \citep{jeffreys1998}.
However, as shown in Fig. \ref{fig:evidence}(b), $9$ input sources have all been successfully identified, while the $10$th source is extremely scattered around the sky, indicating that the data set is best described by the $9$-source model.
The Bayesian evidence does not penalize the model including the excess source as much as we expected, probably because all parameters except the amplitude $\zeta$ of the excess source are unconstrained by the data \citep{taylor2021}.
Therefore, to include a new source in the model, we may require the evidence ratio to be larger than $100$ to draw a decisive conclusion.

We also note that a maximum-likelihood-based method has been developed to resolve multiple CB-SMBHs with PTAs \citep{babak2012,petiteau2013}.
The algorithm neglects the contribution from pulsar terms and consequently discards the pulsar phase parameters.
The likelihood function is then marginalized over inclinations, polarization angles, initial phases and amplitudes of sources and so only depends on locations and frequencies of sources.
As the dimension of the parameter space is reduced to $3N_{\rm s}$, a genetic algorithm can be applied to locate the maximum of likelihood efficiently.
The number of sources in the signal model is also increased gradually until the likelihood can not be improved significantly any more.
\cite{becsy2020} also discards pulsar terms in their signal model but treats the number of GW sources as a free parameter.
The variable-dimension parameter space is sampled by a trans-dimensional MCMC sampler \texttt{BayesHopper} and the number of sources are determined by Bayes factors automatically.
A possible stochastic GW background can also be integrated into the model to search for stand-out sources and unresolved background jointly.
We may first use these methods with simplified signal models to determine the number of sources and estimate their parameters roughly.
Then we can apply DNest method with full signal models to sample the probability distribution of model parameters accurately.

\subsection{Computational complexity}
The time consumed by DNest $t_{\rm tot}$ mainly depends on three factors:
(1) the time of calculating likelihood for given parameters $t_{\rm lh}$;
(2) the length of the Markov chain when sampling the restricted prior distribution $N_{\rm mc}$;
(3) the total number of particles saved in the process $N_{\rm save}$.
Roughly, we have $t_{\rm tot} \approx N_{\rm save} N_{\rm mc} t_{\rm lh}$.

Firstly, $t_{\rm lh}$ is proportional to the number of sources in the model $N_{\rm s}$ and the number of pulsars in the array $N_{\rm p}$.
Secondly, $N_{\rm mc}$ is proportional to the number of free parameters in the model, i.e., $N_{\rm s}(N_{\rm p}+7)$. When the number of pulsars is large, the number of free parameters can be approximated as $N_{\rm s}N_{\rm p}$.
The lower limit of $N_{\rm save}$ depends on the number of levels and the speed of creating levels, and is not easy to estimate.
During the process of DNest, parameter points are ``compressed'' to higher levels gradually until they reach those levels which contain most of the posterior probability.
We define the ratio of the volume of the prior space to the volume occupied by the posterior sample as the compression rate.
Thus the minimum number of levels equals the logarithm of the compression rate, and so proportional to the dimension of the parameter space at first sight.
However, the compression rate contributed by different parameters vary a lot, making the situation more complex.
For example, though the number of pulsar phase parameters surpass that of source parameters largely, the compression rate contributed by them are comparable, since source parameters can be constrained with much smaller uncertainties.
Further more, the speed of creating levels may also change.
We also note that when we update the phase of a pulsar during sampling with FP method, only the timing residual of this pulsar needs to be recalculated.
Similarly, we can record the timing residuals generated by each source independently and update the residuals of a source only when its associated parameters are being updated.
As a result, $t_{\rm lh}$ will not increase a lot when $N_{\rm p}$ and $N_{\rm s}$ increase.

\begin{figure}
    \centering
    \includegraphics[width=\textwidth]{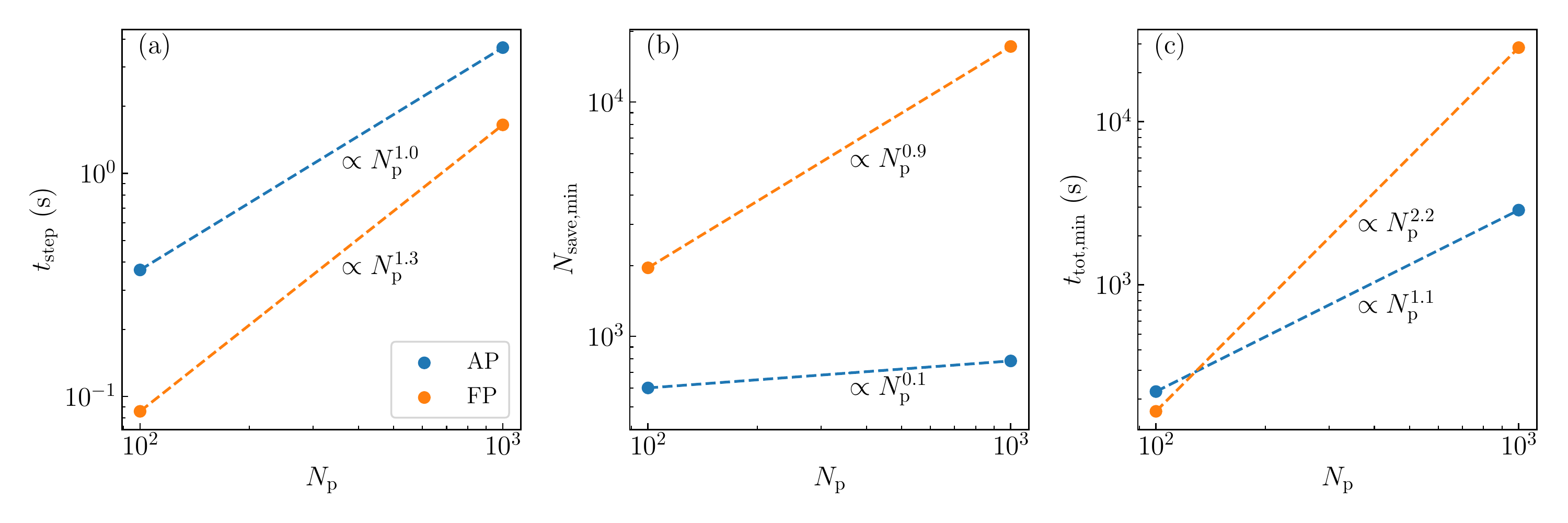}
    \caption{The time consumed by DNest algorithm as the number of pulsars $N_{\rm p}$ increases. 
    (a) The time consumed by saving one sample $t_{\rm step}$, which is the product of the time to calculate likelihood $t_{\rm lh}$ and the length of the Markov chain $N_{\rm mc}$ for sampling the restricted prior ($t_{\rm step} \equiv N_{\rm mc} t_{\rm lh}$).
    For AP method, it increases linearly with $N_{\rm p}$ as the time of calculating likelihood is proportional to $N_{\rm p}$. 
    For FP method, the number of free parameters is proportional to $N_{\rm p}$, while the time of calculating likelihood increases slightly with $N_{\rm p}$, leading to a super-linear dependence on $N_{\rm p}$.
    (b) The minimum number of saved samples to build enough levels and locate the region with maximum of posterior distribution.
    This number increases very slowly with $N_{\rm p}$ for AP method as uncertainties of parameters decrease gradually with $N_{\rm p}$, but it increases nearly linearly with $N_{\rm p}$ for FP method since the dimension of the parameter space is proportional to $N_{\rm p}$.
    (c) The minimum time needed by the algorithm, which is the product of the time consumed by saving one sample and the minimum number of saved samples.
    It increases linearly and quadratically with $N_{\rm p}$ for AP and FP method respectively.
    The algorithm was tested on two chips of Intel Xeon CPU E5-2690 v4 (all of the 28 cores are used).
    \label{fig:time}}
\end{figure}

When $N_{\rm p}$ changes, we can conclude that $t_{\rm tot} \propto N_{\rm save} N_{\rm p}$ for FP, MP and AP methods.
The lower limit of $N_{\rm save}$ increases slowly with $N_{\rm p}$ for MP and AP method because uncertainties of the source parameters are decreased as the SNR of the source increases with $N_{\rm p}$.
As for FP method, the lower limit of $N_{\rm save}$ increases more quickly since the dimension of parameter space also increases with $N_{\rm p}$.
In order to find out the actual dependence of the total time on $N_{\rm p}$, we increase the number of pulsars to $1000$ and redo the calculations.
The result is shown in Fig. \ref{fig:time}.
For AP method, we have $t_{\rm step} \equiv t_{\rm lh} N_{\rm mc} \propto N_{\rm p}$ and $N_{\rm save, min} \propto N_{\rm p}^{0.1}$, leading to $t_{\rm tot, min} \propto N_{\rm p}^{1.1}$.
For FP method, we have $t_{\rm step} \propto N_{\rm p}^{1.3}$ and $N_{\rm save, min} \propto N_{\rm p}^{0.9}$, resulting in $t_{\rm tot, min} \propto N_{\rm p}^{2.2}$.

When $N_{\rm s}$ changes, we have $t_{\rm tot} \propto N_{\rm save} N_{\rm s}$ for FP method.
The lower limit of $N_{\rm save}$ also increases with $N_{\rm s}$ since the dimension of parameter space is proportional to $N_{\rm s}$.
Therefore, the computational complexity is approximately $\order{N_{\rm s}^2}$.
The computation is affordable as long as $N_{\rm s}$ is about $10$ to $20$.
For larger $N_{\rm s}$, we can conduct a hierarchical search as in previous subsection.
If the ranges of parameters of previously found sources can be restricted, the computational complexity for the model with $N_{\rm s}$ sources is closer to $\order{N_{\rm s}}$ rather than $\order{N_{\rm s}^2}$, since the information from previous runs expedites the level building.

\section{Conclusion}
In this paper, we conduct a mock data analysis to test the performance of DNest method in searching for continuous GW signals in pulsar timing residuals.
For a PTA containing $10^2$ pulsars, a Bayesian framework with DNest can overcome the issues of high dimension caused by pulsar phases and performs as well as that with marginalization technique in light of accuracy, robustness and efficiency in search of single sources.
The probability distribution of model parameters can still be sampled effectively even if the number of pulsars increases to $10^3$.
The method can also be used to search for multiple sources simultaneously.
Several strongest sources across a wide range of locations and frequencies can be successfully identified with the method.
The method dose not depend on the analytical form of the signal model and therefore can be generalized to CB-SMBHs with evolving and elliptical orbits or even more complicated models.

\begin{acknowledgments}
We are grateful to the members of the IHEP AGN group and HUST GW astrophysics group for enlightening discussions.
JMW thanks the support by National Key R\&D Program of China through grant -2016YFA0400701, by NSFC through grants NSFC-11991050, -11991054, -11833008, -11690024, and by grant No. QYZDJ-SSW-SLH007 and No.XDB23010400. 
Y.W. is supported by the National Natural Science Foundation of China (NSFC) under Grants No. 11973024 and No. 11690021, and Guangdong Major Project of Basic and Applied Basic Research (Grant No. 2019B030302001). 
\end{acknowledgments}

\appendix

\section{Pulsar timing residuals induced by GW generated by CB-SMBH \label{sec:model}}
\begin{figure}
    \plotone{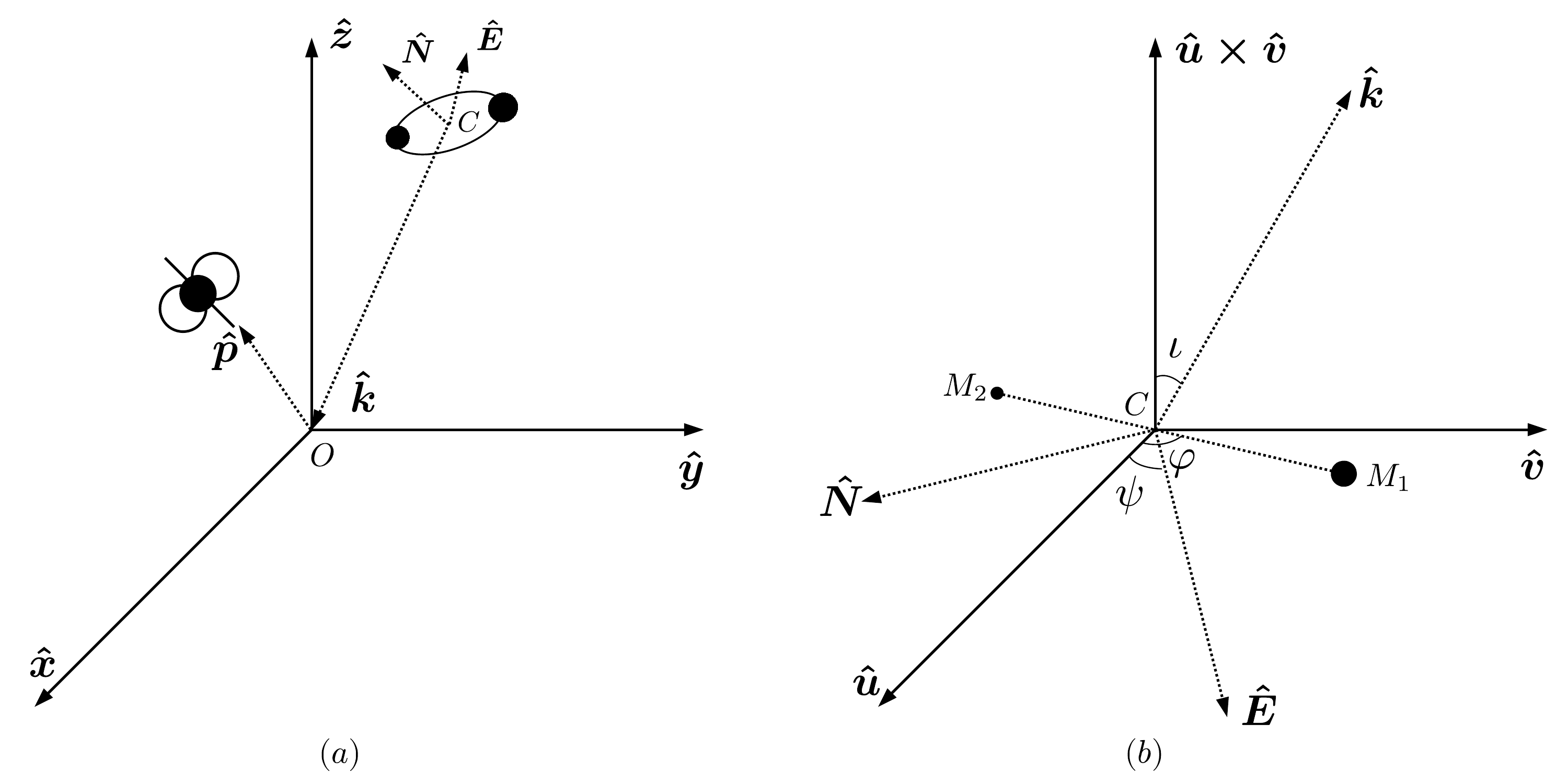}
    \caption{Coordinate systems for the CB-SMBH and pulsar.
    (a) An inertial frame centered at the observer $O$.
    The $z$-axis points to the north celestial pole and $x$-axis to the equinox.
    $C$ is the location of the center of mass of the CB-SMBH and $\vu*{k}$ is the unit vector pointing from $C$ to $O$.
    $\vu*{E}$ and $\vu*{N}$ are the basis vectors of the right ascension and declination of the equatorial coordinates.
    Unit vectors $\vu*{k}$, $\vu*{E}$ and $\vu*{N}$ are perpendicular with each other.
    $\vu*{p}$ is the unit vector pointing from the observer to the pulsar.
    (b) An inertial frame centered at the center of mass of the CB-SMBH $C$.
    $M_1$ and $M_2$ are the primary and secondary black hole respectively.
    The unit vector $\vu*{u}$ lies in the orbital plane and perpendicular to $\vu*{k}$.
    The unit vector $\vu*{v}$ also lies in the orbital plane and perpendicular to $\vu*{u}$.
    The inclination angle $\iota$ is defined as the angle between $\vu*{k}$ and $\vu*{u} \cp \vu*{v}$.
    The polarization angle $\psi$ is defined as the angle between $\vu*{u}$ and $\vu*{E}$.
    The phase angle $\phi$ is defined as the angle between $\vu*{u}$ and $CM_1$.
    \label{fig:BBH}}
\end{figure}

We firstly derive GWs emitted by a binary black hole, following the approach of \cite{wahlquist1987}.
We establish an inertial frame centered at the observer, with $z$-axis pointing to the north celestial pole and $x$-axis to the equinox, as shown in Fig. \ref{fig:BBH}(a).
For a CB-SMBH with right accession $\alpha$, declination $\delta$ and distance $D$, the unit vector pointing from the center of mass of the CB-SMBH to the observer is 
\begin{equation}
    \vu*{k} = -\cos\delta\cos\alpha \vu*{x} - \cos\delta\sin\alpha \vu*{y} - \sin\delta\vu*{z}.
\end{equation}
We choose
\begin{equation}
    \vu*{E} = -\sin\alpha\vu*{x} + \cos\alpha\vu*{y} \qq{and}
    \vu*{N} = -\sin\delta\cos\alpha\vu*{x} - \sin\delta\sin\alpha\vu*{y} + \cos\delta\vu*{z}
\end{equation}
as two orthogonal normalized basic vectors in the plane perpendicular to $\vu*{k}$.
$\vu*{E}$ and $\vu*{N}$ point to the direction of increasing right ascension and declination respectively.

Next, we establish an inertial frame centered at the center of mass of the CB-SMBH, as shown in Fig. \ref{fig:BBH}(b).
We assume that the intersection of the orbital plane and the celestial sphere is
\begin{equation}
    \vu*{u} = \cos{\psi}\vu*{E} + \sin{\psi}\vu*{N},
\end{equation}
and the angle between them is $\iota$.
So the orthogonal to the line of nodes in the orbital plane is
\begin{equation}
    \vu*{v} = \cos \iota (\sin\psi\vu*{E} - \cos{\psi}\vu*{N}) + \sin \iota \vu*{k}.
\end{equation}
Note that $(\vu*{u} \cp \vu*{v})\vdot\vu*{k} = \cos\iota$.

Assuming a non-evolving circular orbit for simplicity, displacements of the primary and secondary black hole in Newtonian limit are
\begin{equation}
    \bm{r}_1 = -D \vu*{k} + \frac{m_2}{m_1+m_2} \bm{r} \qc \bm{r}_2 = -D \vu*{k} - \frac{m_1}{m_1+m_2} \bm{r},
\end{equation}
where $\bm{r} = a [\cos\varphi \vu*{u} + \sin\varphi \vu*{v}]$ is the displacement of the primary black hole relative to the secondary, $\varphi = \omega_{\rm r} t+\varphi_0$ is the orbital phase, $m_1$ and $m_2$ are mass of the two black holes, $a$ is their separation, and $\omega_{\rm r} = [G(m_1+m_2)]^{1/2}a^{-3/2}$ is the angular velocity of the binary in the rest frame.
The quadrupole moment for the system is therefore
\begin{equation}
    \bm{Q} = m_1 \bm{r}_1 \otimes \bm{r}_1 + m_2 \bm{r}_2 \otimes \bm{r}_2
    = MD^2 \vu*{k}\otimes\vu*{k} + \mu \bm{r}\otimes\bm{r},
\end{equation}
where $M=m_1+m_2$ and $\mu=m_1m_2/(m_1+m_2)$ are total mass and reduced mass respectively.

Considering the GW generated by the binary and propagating in the direction $\vu*{k}$, we may use the projection tensor $\bm{I} - \vu*{k} \otimes \vu*{k}$ to obtain the transverse components of $\bm{Q}$,
where $\bm{I}$ is the identity tensor:
\begin{equation}
    \bm{Q}^{\rm T} \equiv (\bm{I} - \vu*{k} \otimes \vu*{k}) \vdot \bm{Q} \vdot (\bm{I} - \vu*{k} \otimes \vu*{k}) = \mu\bm{r}^{\rm T} \otimes \bm{r}^{\rm T},
\end{equation}
where
\begin{align}
    \bm{r}^{\rm T} 
    = (\bm{I} - \vu*{k} \otimes \vu*{k}) \vdot \bm{r} 
    &= a(\cos\varphi\cos\psi + \cos\iota \sin\varphi\sin\psi)\vu*{E} \nonumber \\
    &+ a(\cos\varphi\sin\psi - \cos\iota \sin\varphi\cos\psi)\vu*{N}.
\end{align}
Subtracting the trace then gives $\bm{Q}^{\rm TT}$ as
\begin{equation}
    \bm{Q}^{\rm TT}
    = \mu\left[\bm{r}^{\rm T} \otimes \bm{r}^{\rm T}  - \frac{1}{2}(\bm{r}^{\rm T} \vdot \bm{r}^{\rm T})(\bm{I} - \vu*{k} \otimes \vu*{k}) \right]= Q_+ \bm{e}_+ + Q_{\times} \bm{e}_{\times},
\end{equation}
where
\begin{align}
    Q_+ = \frac{1}{2}\mu a^2 [\cos2\psi(\cos^2\varphi - \cos^2\iota \sin^2 \varphi)  + \sin2\psi \cos\iota \sin 2\varphi], \\
    Q_{\times} = \frac{1}{2}\mu a^2 [\sin2\psi(\cos^2\varphi - \cos^2\iota \sin^2 \varphi) - \cos2\psi \cos\iota \sin 2\varphi ], 
\end{align}
and
\begin{equation}
    \bm{e}_+ = \vu*{E}\otimes\vu*{E} - \vu*{N}\otimes\vu*{N} \qc 
    \bm{e}_{\times} = \vu*{E}\otimes\vu*{N} + \vu*{N}\otimes\vu*{E}
\end{equation}
The GW in transverse traceless gauge (TT gauge) is therefore \citep{misner1973}
\begin{equation}\label{eq:gw}
    \bm{h}^{\rm TT}(t,\bm{x}) = \frac{2G}{c^4D} \eval{\ddot{\bm{Q}}^{\rm TT}}_{t^{\prime}} = h_+ \bm{e}_+ + h_{\times} \bm{e}_{\times},
\end{equation}
with
\begin{align}\label{eq:gw_comp}
    h_+ = \frac{2 (G\mathcal{M}_{\rm r})^{5/3} \omega_{\rm r}^{2/3}}{c^4D} [-\cos2\psi(1+\cos^2\iota)\cos2\varphi - 2\sin2\psi \cos\iota \sin 2\varphi],
    \nonumber \\
    h_{\times} = \frac{2 (G\mathcal{M}_{\rm r})^{5/3} \omega_{\rm r}^{2/3}}{c^4D} [-\sin2\psi(1+\cos^2\iota)\cos2\varphi + 2\cos2\psi \cos\iota \sin 2\varphi].
\end{align}
Here, $\mathcal{M}_{\rm r} \equiv (m_1m_2)^{3/5}/(m_1+m_2)^{1/5}$ is the rest-frame chirp mass and $t^{\prime} = t - |\bm{x}+\vu*{k}D|/c$ is the retarded time.

If we take the expansion of universe into consideration and assume that the redshift of the CB-SMBH is $z$, Eq. (\ref{eq:gw_comp})
should be revised to
\begin{align}
    h_+ = \frac{2 (G\mathcal{M})^{5/3} \omega^{2/3}}{c^4D_{\rm L}} [-\cos2\psi(1+\cos^2\iota)\cos2\varphi - 2\sin2\psi \cos\iota \sin 2\varphi],
    \nonumber \\
    h_{\times} = \frac{2 (G\mathcal{M})^{5/3} \omega^{2/3}}{c^4D_{\rm L}} [-\sin2\psi(1+\cos^2\iota)\cos2\varphi + 2\cos2\psi \cos\iota \sin 2\varphi],
\end{align}
where $D_{\rm L}$ is the luminosity distance of the CB-SMBH, $\mathcal{M} = (1+z)\mathcal{M}_{\rm r}$ is the redshifted chirp mass, $\omega = (1+z)^{-1}\omega_{\rm r}$ is the observed orbital angular frequency, $\varphi = \varphi_0 + \omega t$ is the observed orbital phase \citep{holz2005}.

If we further consider the evolution of the binary orbit due to the radiation of GWs, the merger time in the observer's frame is
\begin{equation}
    t_{\rm merge} = \frac{5c^5}{256} (G\mathcal{M})^{-5/3} \omega^{-8/3} = \num{4.4e4} \left(\frac{\mathcal{M}}{\num{e9}M_{\sun}}\right)^{-5/3} \left(\frac{\mathcal{\omega}}{\SI{e-8}{\hertz}}\right)^{-8/3} \si{yr}.
\end{equation}
On such a time scale, the angular velocity of the binary is no longer constant but varies with time as
\begin{equation}
    \omega(t) = \omega_0\left(1-\frac{t}{t_{\rm merge}}\right)^{-3/8},
\end{equation}
where $\omega_0$ is the observed angular frequency at initial time.
The evolution of orbital phase will be modified to
\begin{equation}
    \varphi(t) = \varphi_0 + \int_0^t \omega(t') \dd{t'} = 
    \varphi_0 + \frac{8\omega_0 t_{\rm merge}}{5} \left[1-\left(1-\frac{t}{t_{\rm merge}}\right)^{5/8}\right],
\end{equation}
which reduces to $\varphi(t) = \varphi_0 + \omega_0 t$ when $|t| \ll t_{\rm merge}$.

To derive the impact of gravitational waves on the propagation of radio pulses from pulsars, we follow the approach in \citet{estabrook1975}.
As the wavelength of the GW is much shorter than the distance between the observer and pulsars in the Galaxy, which is further shorter than the distance of the CB-SMBH, the GW can be well approximated by a plane wave propagating in the direction $\vu*{k}$. 
So the space-time possesses three Killing vectors:
\begin{equation}
    \xi_a = (0, \vu*{E}) \qc \xi_b = (0, \vu*{N}) \qq{and} \xi_c = (1, \vu*{k}).
\end{equation}
Now, suppose there is a pulsar in direction $\vu*{p}$, emitting a photon towards us. 
The frequency of the photon is $\nu$ in the rest frame of the pulsar. 
Thus, the four wave vector of the photon to $\order{h}$ is
\begin{equation}
    k^{\mu} = \nu \left(1, -\left(\delta_{ij} - \frac{1}{2}h_{ij}\right)\hat{p}_j\right),
\end{equation} 
or as a covariant vector
\begin{equation}
    k_{\mu} = -\nu \left(1, \left(\delta_{ij} + \frac{1}{2}h_{ij}\right)\hat{p}_j\right).    
\end{equation}

Due to the properties of Killing vectors and geodesics, the quantities $k_{\mu}\xi^{\mu}$ must be conserved during the propagation of the photon. 
When the photon reaches us, the frequency has shifted to $\nu + \var{\nu}$ and direction to $\vu*{p} + \var{\vu*{p}}$. 
We have
\begin{align}\label{eq:Killing}
    &\nu \left(\hat{p}_i \hat{E}_i + \frac{1}{2}h_{ij}(x)\hat{E}_i\hat{p}_j\right) =  (\nu+\var{\nu}) \left((\hat{p}_i + \var{\hat{p}_i}) \hat{E}_i + \frac{1}{2}h_{ij}(y)\hat{E}_i(\hat{p}_j + \var{\hat{p}_j})\right) 
    \nonumber \\
    &\nu \left(\hat{p}_i \hat{N}_i + \frac{1}{2}h_{ij}(x)\hat{N}_i\hat{p}_j\right) =  (\nu+\var{\nu}) \left((\hat{p}_i + \var{\hat{p}_i}) \hat{N}_i + \frac{1}{2}h_{ij}(y)\hat{N}_i(\hat{p}_j + \var{\hat{p}_j})\right)
    \nonumber \\
    &\nu \left(1+\hat{k}_i\hat{p}_i\right) = (\nu+\var{\nu})\left(1+\hat{k}_i(\hat{p}_i + \var{\hat{p}_i})\right)
    \nonumber \\
    &(\hat{p}_i + \var{\hat{p}_i})(\hat{p}_j + \var{\hat{p}_j}) = 1,
\end{align}
where $x$ and $y$ are space-time coordinates of when the photon is emitted and received respectively.

To the first order of $h$, the solution to Eq. (\ref{eq:Killing}) gives
\begin{equation}
    \frac{\var{\nu}}{\nu}
    = F^+ \Delta h_+ + F^{\times}\Delta h_{\times},
\end{equation}
where
\begin{equation}
    F^+ = \frac{1}{2} \frac{(\vu*{N}\vdot\vu*{p})^2 - (\vu*{E}\vdot\vu*{p})^2}{1 + \vu*{p}\vdot\vu*{k}} \qc
    F^{\times} = -\frac{(\vu*{E}\vdot\vu*{p})(\vu*{N}\vdot\vu*{p})}{1 + \vu*{p}\vdot\vu*{k}},
\end{equation}
and
\begin{equation}
    \Delta h_{+,\times} =  h_{+,\times}(t,0) - h_{+,\times}(t - d_{\rm p}/c, d_{\rm p}\vu*{p}) = h_{+,\times}(t,0) - h_{+,\times}(t - d_{\rm p}(1+\vu*{p}\vdot\vu*{k})/c,0).
\end{equation}
Here, $d_{\rm p}$ is the distance of the pulsar.
We also define the pulsar lag $\tau \equiv d_{\rm p}(1+\vu*{p}\vdot\vu*{k})/c$ and pulsar time $t_{\rm p} \equiv t - \tau$ respectively.

The timing residual induced by the GW for an observer at Earth is
\begin{equation}
    s(t) = -\int \frac{\var{\nu}}{\nu} \dd{t} = F^+ \Delta s_+ + F^{\times}\Delta s_{\times}
\end{equation}
where
\begin{align}
    \Delta s_{+,\times} &= s_{+,\times}(t) -  s_{+,\times}(t_{\rm p}), \nonumber \\
    s_{+}(t) &= \frac{(G\mathcal{M})^{5/3}}{c^4D_{\rm L}\omega(t)^{1/3}} [\cos2\psi(1+\cos^2\iota)\sin2\varphi(t) - 2\sin2\psi \cos\iota \cos 2\varphi(t)], \nonumber \\
    s_{\times}(t) &= \frac{(G\mathcal{M})^{5/3}}{c^4D_{\rm L}\omega(t)^{1/3}} [\sin2\psi(1+\cos^2\iota)\sin2\varphi(t) + 2\cos2\psi \cos\iota \cos 2\varphi(t)],
\end{align}
For the pulsar term $s_{+,\times}(t_{\rm p})$, we also note that
\begin{equation}
    \omega(t_{\rm p}) = \omega_{\rm p} \left(1 - \frac{t}{t_{\rm merge} + \tau}\right)^{-3/8} \qc
    \varphi(t_{\rm p}) = \varphi_{\rm p} + \frac{8\omega_{\rm p}(t_{\rm merge} + \tau)}{5} \left[1-\left(1-\frac{t}{t_{\rm merge}+\tau}\right)^{5/8}\right].
\end{equation}
where $\omega_{\rm p} \equiv \omega(-\tau)$ and $\varphi_{\rm p} \equiv \varphi(-\tau)$.

\section{Evolution of binary orbits \label{sec:evolution}}
We have neglected evolution of binary orbits in our mock data analysis.
However, the detectable CB-SMBH usually has a chirp mass larger than $10^9 M_{\sun}$ and an orbital period less than $10$ years, and thereby its merger time will be comparable to light travel times between the Earth and pulsars, making frequencies of the Earth term and pulsar term different.
Fortunately, our method can be generalized to the case of evolving orbits directly by further including merger times of sources and distances of pulsars in the model.
For simplicity, we assume relative uncertainties of pulsar distances given by electromagnetic observations are $20\%$ and using gaussian priors for them.
The total dimension of the parameter space will be $N_{\rm s}(8 + N_{\rm p}) + N_{\rm p}$.

\begin{figure}
    \centering
    \includegraphics[width=\textwidth]{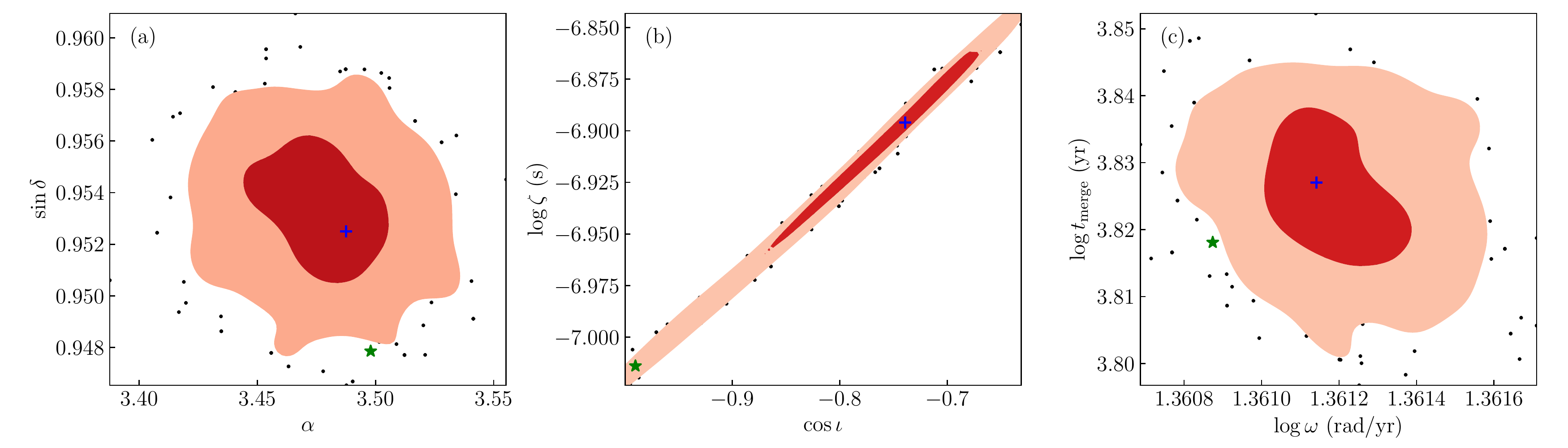}
    \caption{Posterior probability distributions of some model parameters of the GW source with an evolving orbit.
    (a) Distributions of the right accession and declination of the source.
    The red and pink colors represent the $1\sigma$ and $2\sigma$ regions of the distribution.
    Black dots are posterior samples outside the $2\sigma$ region.
    The green star marks the input parameters, while the blue cross indicates the peak of the distribution.
    (b) Distributions of the inclination and amplitude of the source.
    (c) Distributions of the angular frequency and merger time of the source.
    \label{fig:evolve}}
\end{figure}

We note that merger times of a fair fraction of CB-SMBHs generated by the method in subsection \ref{subsec:simu} are quite short ($\lesssim \SI{10}{yr}$), and so they are hardly caught by a typical PTA program.
Therefore, we draw amplitude of timing residuals $\zeta$ and merger time $t_{\rm mrege}$ rather than chirp mass and luminosity distance log-uniformly when generating mock populations of CB-SMBHs.
The ranges of $\zeta$ and $t_{\rm mrege}$ are $[\num{e-15},\num{e-7}]\si{\second}$ and $[\num{e2},\num{e8}]\si{yr}$ respectively.
Then we perform a blind search for a single target via DNest method, as done in subsection \ref{subsec:blind_single}.
The result is shown in Fig. \ref{fig:evolve}.

The source with the highest SNR ($\rho \sim 120$) is successfully identified.
Fig. \ref{fig:evolve}(a) shows the distribution of right ascension and declination of the source.
The actually location of the source lies near the $2\sigma$ region of the posterior sample.
The $2\sigma$ region of the posterior sample has a size of $\sim \SI{3}{deg^2}$, and the angular distance between the source and distribution center is about $\SI{0.7}{\degree}$.
The second panel shows the distribution of the amplitude of timing residuals and the inclination of the orbital plane.
The two parameters is still highly correlated, and the uncertainty of amplitude is about $\SI{0.1}{dex}$.
The third panel displays the distribution of the angular frequency of the GW and merger time of the CB-SMBH, which are weakly correlated.
The frequency can still be constrained precisely.
Though the merger time is much more longer than the time span of the PTA program, it can be determined quite accurately by detecting the frequency difference between the Earth term and pulsar terms.
This shows the potential of CB-SMBHs with $t_{\rm merge} \lesssim \SI{e4}{yr}$ as standard sirens for cosmology if their electromagnetic counterparts can be identified.
In a word, DNest also performs well when evolution of binary orbits is taken into account.

\newcommand{\noop}[1]{}

\end{document}